\documentclass[aip, pof, preprint]{revtex4-1}
\usepackage{epsfig,graphics,subfigure,amssymb,amsmath,subeqnarray,color,xcolor}
\def\F{\mathbf{F}}
\def\L{\mathbf{L}}
\def\U{\mathbf{U}}
\def\e{\mathbf{e}}
\def\D{\mathbf{D}}

\def\Db{\mathbf{Db}}

\def\B{\mathbf{B}}
\def\t{\mathbf{t}}
\def\0{\mathbf{0}}

\def\bOmega{\boldsymbol{\Omega}}
\newcommand{\degree}{\ensuremath{^\circ}}
\newcommand{\upl}{\ensuremath{^{(0)}}}
\newcommand{\upf}{\ensuremath{^{(1)}}}
\newcommand{\ups}{\ensuremath{^{(2)}}}
\let\originaleqref=\eqref
\renewcommand{\eqref}{Eq.~\originaleqref}

\begin{document}

\title{Wobbling-to-swimming transition i}
\title{The wobbling-to-swimming transition of rotated helices}
\author{Yi Man and Eric Lauga}
\email{e.lauga@damtp.cam.ac.uk}
\affiliation{Department of Applied Mathematics and Theoretical Physics, 
University of Cambridge,
Wilberforce Road, Cambridge CB3 0WA, UK.}
\date{\today}
\begin{abstract}
A growing body of work aims at designing and testing  micron-scale synthetic swimmers. One method, inspired by the locomotion of flagellated bacteria, consists of applying a rotating magnetic field to a rigid, helically-shaped, propeller attached to a magnetic head. When the resulting device, termed an artificial bacteria flagellum, is aligned perpendicularly to the applied field, the helix rotates and the swimmer moves forward. Experimental investigation of  artificial bacteria flagella shows  that at low  frequency of the applied field, the axis of the helix does not align perpendicularly to the field but wobbles around the helix, with an angle increasing as the inverse of the field frequency. By numerical computations and asymptotic analysis, we provide a theoretical explanation for this wobbling behavior. We numerically demonstrate the wobbling-to-swimming transition as a function of the helix geometry and the dimensionless Mason number which quantifies the ratio of viscous to magnetic torques. We then employ an asymptotic expansion for near-straight helices to derive an analytical estimate for the wobbling angle 
allowing to rationalize our computations and past experimental results. These results can help guide future design of artificial helical swimmers.

\end{abstract}
\maketitle

%%%%%%%%%%%%%%%%%%%%%
\section{Introduction}

A significant effort in the fluid mechanics literature has focused on the individual and collective  dynamics of  low-Reynolds number swimmers. The original work in the field,  started decades ago,  aimed at quantifying the kinematics and energetics of biological microorganisms such as bacteria, spermatozoa, or plankton  \cite{brennen,pedley92}. Recently, fluid mechanical studies have  also focused on  the dynamics of artificial microswimmers, motivated in part by potential applications of small-scale locomotion to targeted drug delivery, micro-surgery, or diagnostics \cite{nelson10,abbot,kosa,Kosa08}.

As is now well known, the physics of swimming in Stokes flows is quite different from that of  swimming on a human length scale. The oft-cited distinguishing property is the scallop theorem \cite{purcell}, which states that locomotion by a sequence of shape which is  reciprocal  (i.e. identical under a time-reversal symmetry)  leads to zero net propulsion. So,  for example,  the flapping of the fins of a scuba diver does not work on small length scales. Biological organisms are able to circumvent the constraints of the theorem by deforming their bodies or appendages (flagella) in a wave-like fashion breaking the time-reversibility requirement \cite{lauga09,lauga11}.

Broadly speaking, three different types of synthetic micro/nano swimmers have been proposed. The first kind exploits chemical reactions on patterned catalytic surfaces and the flow resulting from phoretic or electrokinetic motion \cite{paxton04,golestanian05,howse07,golestanian07,chemical1,chemical2, chemical3}. The second type, actuated by external (typically magnetic) fields, exploits the presence of a nearby surface to escape from the constraint of the scallop theorem under a time-periodic actuation \cite{surface1, surface2, surface3}. The final category of synthetic swimmer is inspired  by  the locomotion strategy of flagellated bacteria, namely the rotation of one or many helical flagella \cite{ecoli}. Flexible and straight filaments can acquire  chirality when actuated in rotation by an external field, leading to propulsion scaling nonlinearly with the field frequency \cite{flexible1,flexible2,OnshunSoftMatter}.  Alternatively, the chirality can be built in the design and fabrication of the filament. The simplest examples are rigid helical filaments attached to magnetic heads which, under an externally rotating magnetic field, rotate as  cork-screws and lead to forward  motion \cite{NelsonRev,GhoshFischer09,ghosh12}.

In this paper we focus on the dynamics of these rigid helical  propellers, referred to in the literature as artificial bacteria flagellum (or flagella). Different experimental protocols have been proposed to design them capable of precise motion control yet  high speed. One method uses a self-scrolling technique to fabricate a  nanobelt-based artificial bacteria flagellum consisting of a helical metal tail attached to a thin square soft-magnetic metal head  \cite{NelsonRev}. This helix has a width of 1.8 $\mu m$, a wavelength of 10 $\mu $m, and is equipped with a square head, of width 1.8 $\mu$m   \cite{NelsonRev, NelsonNonideal}, which can alternatively be replaced  by a microholder to allow cargo transport  \cite{NelsonCargo}.    A different design was implemented using glancing angle deposit \cite{GhoshFischer09}. The helix  in this case is made of silicon dioxide, and has a width of $200-300$ nm  and a length of  $1-2$ $\mu$m  \cite{GhoshFischer09}. In both cases,  the artificial bacteria flagella possess  a magnetic moment perpendicular to the  long axis of the helix and are  controlled by an externally-rotating magnetic field generated by triaxial Helmholtz coils. Under this actuation, the nano-belt based swimmer in Ref.~\cite{NelsonRev} with four wavelengths is able to swim with a velocity  {\color{black}of approximately} 5 $\mu$m/s at an input frequency of about 10 Hz while the glass (silicon dioxide) helix from Ref.~\cite{GhoshFischer09} can  swim at a velocity  {\color{black}of approximately} 40 $\mu$m/s at a field frequency of about 150 Hz.

 {\color{black}When the axis of the helix aligns with the swimming direction, local thrust arising from the fluid drag is everywhere  directed along the helix axis \cite{lauga09}.} Therefore, in order for artificial bacteria flagella to be efficient, it is experimentally important that their axis remain always  perpendicular to applied field, in which case one would then expect a swimming velocity  scaling linearly  with the {\color{black} field} frequency \cite{NelsonRev, NelsonNonideal, GhoshFischer09}.  Experimentally,  problems are however seen to arise at both high and low frequencies. When the frequency is larger than a critical value (step-out frequency), the viscous {\color{black}torque} becomes larger than the typical magnetic torque  and the helix can no longer follow the field \cite{NelsonNonideal,ghosh12}.

\begin{figure}[t]
\includegraphics[width=0.5\textwidth]{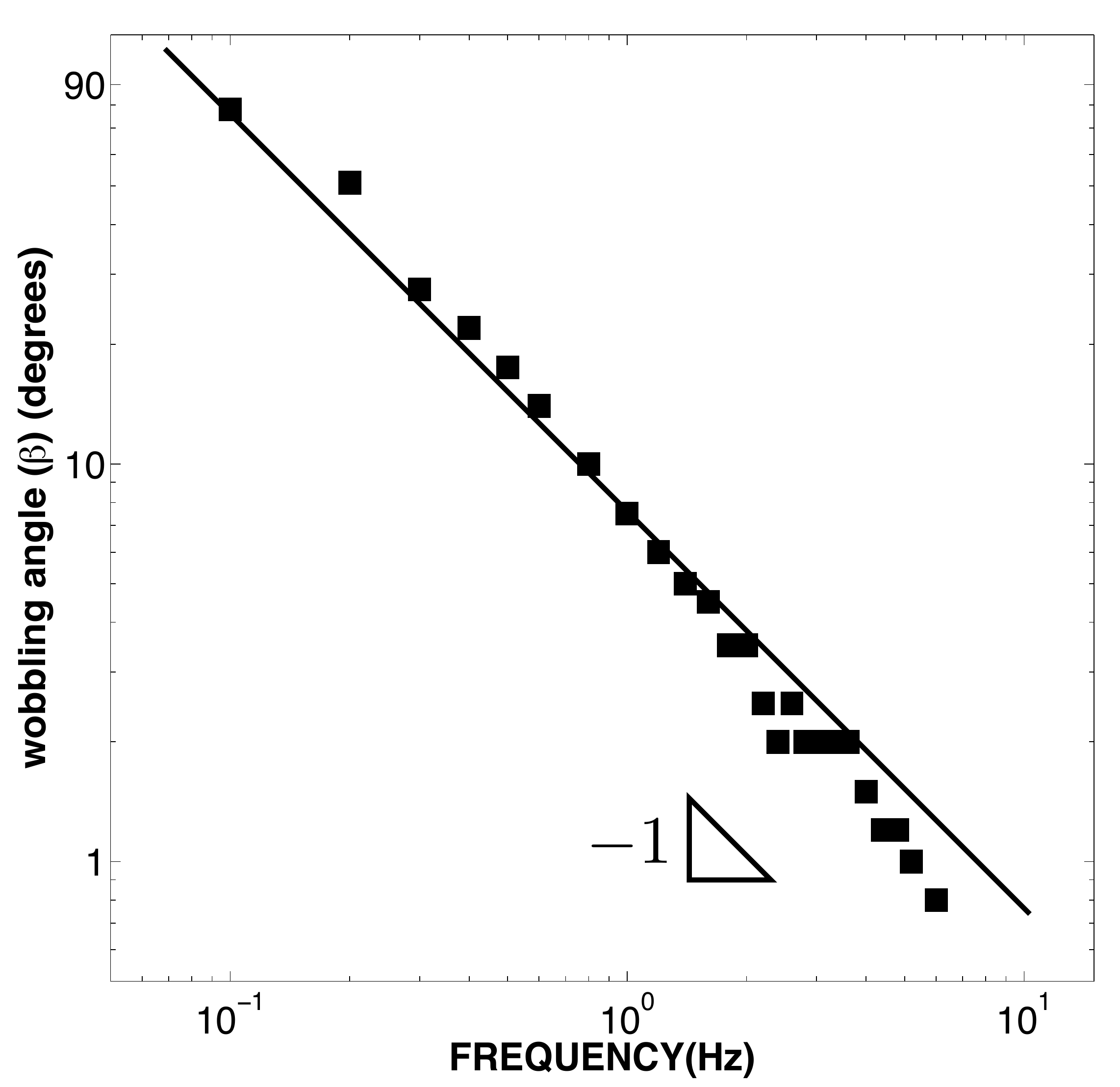}
\caption{Experimental dependence of helix (artificial bacterial flagella) wobbling angle as a function of the magnetic field frequency  (log-log scale), reproduced from Ref.~\cite{NelsonNonideal}. The line in the figure has the slope of $-1$ showing that the wobbling angle scales as the inverse  of the frequency. \label{fig:experimental}}
\end{figure}

Perhaps more surprisingly, at low {\color{black}field} frequency, the axis of the artificial bacteria flagellum is observed experimentally {\color{black}not to remain }perpendicular to the plane of the magnetic field but instead undergoes a periodic {\color{black}precessive} movement at an angle with the (desired) average swimming direction. This movement,  referred to as wobbling, is best characterized by an average  wobbling angle \cite{NelsonNonideal}, with straight swimming  corresponding to a wobbling angle of zero. At {\color{black}low} frequencies (typically below a few Hertz in the experiments of Ref.~\cite{NelsonNonideal}) the wobbling angle is observed to increase as the  frequency decreases, from zero up  to a maximum of  ninety degrees at the lowest test frequency (meaning that, in this limit, the helix axis actually rotates at a right angle with respect to the desired swimming direction). Plotting the measured wobbling angle from Ref.~\cite{NelsonNonideal} in Fig.~\ref{fig:experimental} we see that the wobbling angle, $\beta$, scales as the inverse first power of the field frequency, $\beta \sim \omega^{-1}$. In this paper, we use numerical computations and a theoretical analysis to provide a physical model for this wobbling behavior.

Our paper is organized {\color{black}as} three sections. We first build a mathematical model of the dynamics of artificial bacteria flagella based on the mechanical balance of forces and torques with resistive force theory {\color{black} used to describe} the hydrodynamics. We then employ numerical computations to characterize the steady-state locomotion of artificial bacteria flagella and demonstrate numerically a transition from wobbling to swimming with a similar inverse frequency scaling as the one seen experimentally. We finally employ an asymptotic analysis to provide an analytical model  for the wobbling behavior, recovering in particular the scaling with the frequency and predicting the geometrical factors affecting it.

%%%%%%%%%%%%%%%%%%%%%%%%%%%%%%
\section{Modeling the dynamics of artificial bacterial flagella}

\subsection{Geometry}

\begin{figure}[t]
\includegraphics[width=.7\textwidth]{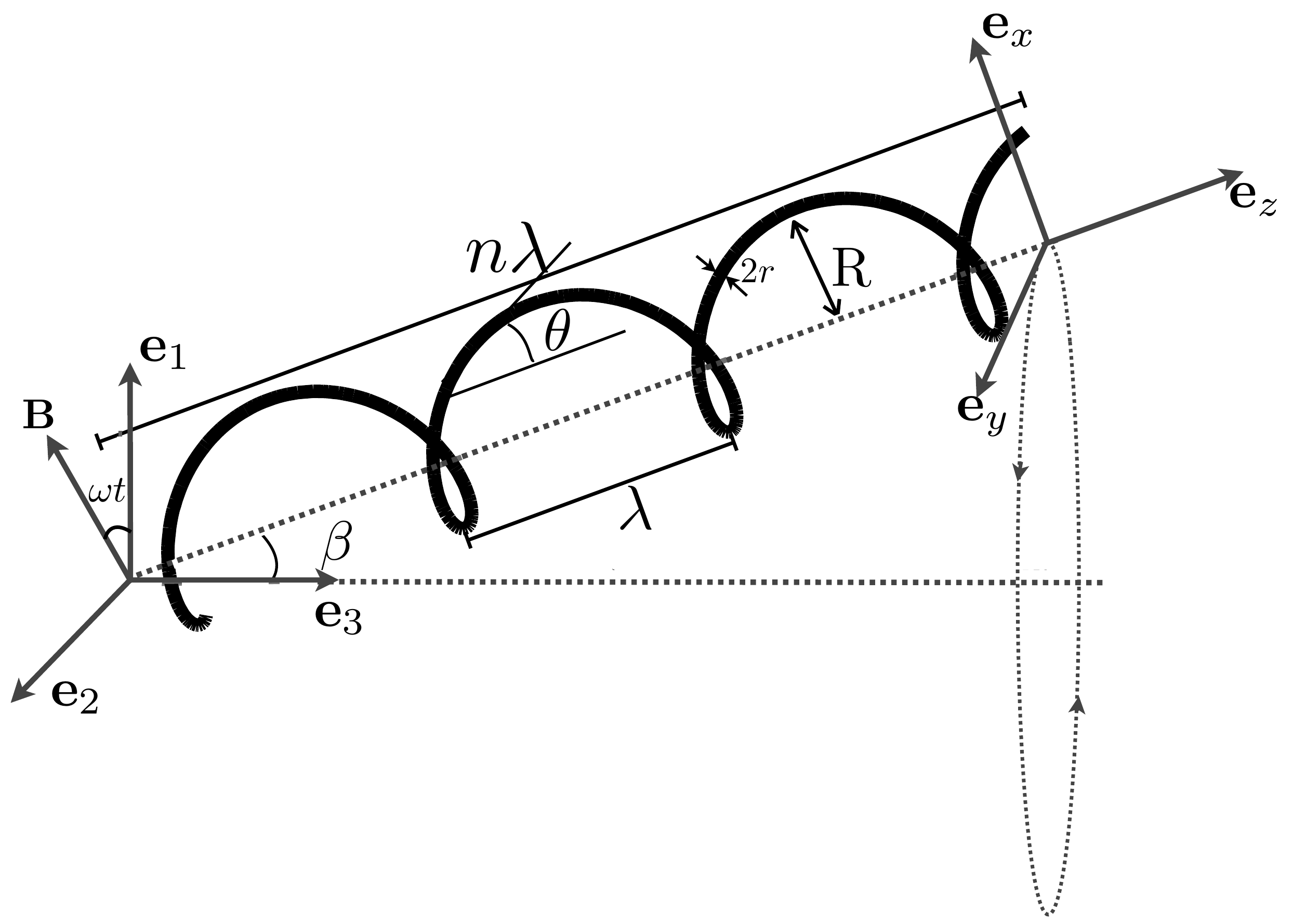}
\caption{Geometry of the rigid helix as a model for an artificial bacterial flagellum. The parameters of the  helix are its  wavelength $\lambda$, the helix angle $\theta$, the radius $R$,  the diameter of the cross-section  $2r$, and the number of wavelengths $n$. The  wobbling angle,   which is the angle between the axis of the helix  and the mean forward velocity, is denoted $\beta$. The vectors $(\e_1, \e_2, \e_3)$  and $(\e_x, \e_y, \e_z)$ constitute the laboratory frame and body frame respectively. The magnetic field, $\B$, rotates in the $(\e_1, \e_2)$ plane with frequency $\omega$. 
\label{fig:geo}}
\end{figure}

We model an artificial bacterial flagellum as a rigid helix  of circular cross-section with no head, as illustrated in Fig.~\ref{fig:geo}. The geometric parameters are therefore: the radius of helical body ($R$), its wavelength along the helix axis ($\lambda$), the helix angle ($\theta$), the radius of the helix  cross-section ($r$) and the number of wavelength ($n$).  We set up two frames of reference, the laboratory frame, denoted $(\e_1,\e_2, \e_3)$,  and the frame attached to the body, termed $(\e_x, \e_y,\e_z)$. The  wobbling angle,   which is the angle between the axis of the helix ($\e_z$) and the desired direction of the forward velocity ($\e_3$), is denoted $\beta$. In the body frame,  the shape of the helix remains constant  and is described by the location vector, $\mathbf{x}_h$, of material points along the helix centerline as
\begin{equation}\label{eq:helix}
[\mathbf{x}_h]_{body}=[R\cos(\kappa\alpha s), R\sin(\kappa\alpha s), \alpha s]^T,
\end{equation}
where $s$ is the arclength coordinate, $\kappa$ is the wavenumber, and $\alpha=\cos\theta$. In this paper, we use the subscripts ``body" and ``lab" when we explicitly give the components of a vector to denote in which frame of reference  these components are to be understood. We also denote vectors as column vectors  and thus use a transpose sign, $^T$, when the coordinates are  written along a row.

\subsection{Dynamics}
\subsubsection{External torque}
The locomotion of the artificial bacteria flagellum  is actuated by an external magnetic field, $\B$, 
rotating in the  $(\e_1, \e_2)$ plane with  frequency  $\omega$ and amplitude  $B_0$ as
\begin{equation}\label{eq:magnetic}
[\B]_{lab}=B_0[ cos(\omega t), sin(\omega t), 0]^T.
\end{equation}
Experimentally  this magnetic field provides an external torque to the head of the artificial flagellum {\color{black}but} no external force.  Since the presence of the head is not necessary {\color{black}from} a hydrodynamic standpoint to obtain wobbling, we ignore it hydrodynamically in our model. As a proxy for the  head's magnetization we assume that a constant magnetic moment of magnitude $H$, is present along the $\e_y$ direction in the body frame. The torque imposed by the magnetic field, $\L_m$, is then obtained as
\begin{equation}\label{eq:magnetic torque}
\L_m= H\e_y\times \B.
\end{equation}

\subsubsection{Hydrodynamics}
In the experiments of Ref.~\cite{NelsonNonideal}, the typical rotation frequency of the field  reaches a maximum of tens of Hz and the helix radius is on the order of a few microns, leading to a typical Reynolds number for locomotion in water much less than unity. The fluid dynamics for the flow around the artificial bacteria flagellum is thus well approximated by a solution to the incompressible Stokes equations. Given the slenderness of the helical geometry, it is {\color{black}convenient} to use resistive-force theory to describe the distribution of forces and moments on the moving helix \cite{RFT,Cox1970,Batchelor1970,lauga09}. The basic assumption of resistive-force theory is that the  hydrodynamic forces acting on the slender helix moving through the fluid per unit length, $d\F_v$, is locally proportional, albeit in an anisotropic fashion, to the relative velocity, $\U$, between the helix and the surrounding fluid. Given the unit tangent vector along the helix, $\t={{\rm d}\mathbf{x}_h}/{{\rm d}s}$, and the shear viscosity of the fluid, $\mu$,  this  linear relationship is written as 
 \begin{equation}\label{eq:RFT_dF}
d\F_v=-\xi_\parallel \mathbf{U_\parallel} - \xi_\perp\mathbf{U_\perp},
\end{equation}
where $\U_\parallel=(\U\cdot\t)\t$ and $\U_\perp=\U-\U_{\parallel}$ are the components of velocity along the tangential and normal directions respectively and  $\xi_\parallel$ and $\xi_\perp$ are the corresponding viscous drag coefficients \cite{RFT} 
\begin{subeqnarray}\label{eq:coefficient}
\xi_{\parallel}&\approx&\frac{2\pi\mu}{\ln({2\lambda}/{r})-{1}/{2}},\\
\xi_{\perp}&\approx&2\xi_{\parallel}.
\end{subeqnarray}
{\color{black}Resistive-force theory} is the leading-order term in a systematic expansion of the flow around slender bodies in powers of $\sim \left(\ln {L}/{r}\right)^{-1}$, where $L$ is the total length of helix  \cite{Cox1970, Batchelor1970,keller76-jfm,johnson80,lauga09}. {\color{black} Although resistive-force theory can lose some features of the interrelations between the fluid and curved geometry\cite{Pak2012,Jung07}, we first apply it for its simplicity and convenience. If the resistive-force theory doesn't work well, we need to consider the expansion with higher orders.}

With the force distribution, $d\F_v$, known everywhere along the helix, it is straightforward to  calculate its contribution to the net moment per unit length acting on the helix as $\mathbf{x}_h\times d\F_v$. An additional contribution to a torque on the helix arises from its instantaneous rotation around its centerline, described by a moment density $d\L_r=4\pi\mu r^2(\bOmega\cdot\mathbf{t})\mathbf{t}$  where $\bOmega$ is the helix rotation rate \cite{Spinning} .  This term is typically of order $\sim ({r}/{L})^2$ smaller than the torque arising from resistive-force theory and can usually be disregarded, but it becomes important when the helix is a near-rod as it prevents  its viscous mobility matrix  to become singular. We therefore keep it in our formulation and write  the net
 hydrodynamic torque per unit length acting on the helix as
\begin{equation}\label{eq:torque_dL}
d\L_v=\mathbf{x}_h\times d\F_v+d\L_r.
\end{equation}

Integrating Eqs.~(\ref{eq:RFT_dF}) and (\ref{eq:torque_dL}) along the flagellum finally leads to a linear relationship relating  the total hydrodynamic force, $\F_v$, and torque, $\L_v$, to the  rigid-body velocity, $\U$,  and rotation rate, $\bOmega$,  of the swimming helix as
\begin{equation}\label{eq:RFT_matrix}
\begin{bmatrix}
\F_v\\
\L_v
\end{bmatrix}=
\D
\begin{bmatrix}
\U\\
\bOmega
\end{bmatrix}\cdot
\end{equation}
The  $6\times6$ viscous resistance tensor, $\D$, has constant coefficients in the body frame of the helix. The calculation for its components is straightforward but tedious, and the final  nondimensionalized results are  given in Appendix \ref{app}.

\subsubsection{Force and torque balance}

The dynamics of the helix is governed  by the balance of force and torque as
\begin{subeqnarray}\label{eq:force_balance}
 \F_v&=&\mathbf{0},\\
 \L_v+\L_m&=&\mathbf{0}.
\end{subeqnarray}
Since the viscous resistance tensor, $\D$, has constant coefficients when expressed  in the body frame, it is necessary to enforce \eqref{eq:force_balance} in the body frame. The {\color{black}kinematics} of the body frame is  described by the three vector equations
\begin{equation}\label{eq:ODEs}
\frac{d\e_x}{dt}=\bOmega\times\e_x,\,
\frac{d\e_y}{dt}=\bOmega\times\e_y,\,
\frac{d\e_z}{dt}=\bOmega\times\e_z.
\end{equation}
The combination of {\color{black}Eqs.~ (\ref{eq:RFT_matrix}), (\ref{eq:force_balance}) and (\ref{eq:ODEs})}has a total of 15 unknowns (6 kinematics components and 9 components of the rotating frame coordinates) together with a $6\times6$ linear system (Eqs.~\ref{eq:RFT_matrix}, \ref{eq:force_balance}) and a $9\times9$ ODE system (Eq.~\ref{eq:ODEs}) leading to a closed system.

\subsection{Nondimensionalization}
In order to nondimensionalize the problem we use the  wavelength $\Lambda$ calculated along the helix centerline as length scale  ($\Lambda={\lambda}/{\cos\theta}$) and the inverse of magnetic field frequency,  $\omega^{-1}$, as the characteristic time scale. Geometrically, there are three independent dimensionless parameters describing the helix, namely the helix angle $\theta$, the  number of wavelengths $n$, and the dimensionless radius of the flagellum, which we denote $\gamma$. The viscous drag coefficients are nondimensionalized by the fluid viscosity and thus we have, using bars to denote dimensionless quantities,
\begin{equation}
\begin{split}
\bar{\xi}_\parallel&=\frac{\xi_\parallel}{\mu}=\frac{2\pi}{\ln({2\cos\theta}/{\gamma})-{1}/{2}},\\
\bar{\xi}_\perp&=\frac{\xi_\perp}{\mu} =2\bar{\xi}_{\parallel}.
\end{split}
\end{equation}
Using $B_0$ as the scale of the magnetic field we have therefore  $[\bar{\B}]_{lab}=[ \cos\bar{t}, \sin\bar{t}, 0]^T$, where $\bar{t}$ is the dimensionless time, $\bar{t}=\omega t$. The dimensionless version of the force and torque balance,  \eqref{eq:force_balance}, is given by
\begin{subeqnarray}
\label{eq:bar_force_balance}
\mu\omega\Lambda^2  \bar{\F}_v&=&\mathbf{0},\\
\displaystyle \frac{\mu\omega\Lambda^3}{ HB_0}\bar{\L}_v+\bar{\L}_m&=&\mathbf{0}.\slabel{last}
\end{subeqnarray}
Inspecting \eqref{last} we observe that   a dimensionless group appears in the torque balance.  It is termed a Mason number, ${\rm Ma}={\mu\omega\Lambda^3}/{ HB_0}$, and quantifies the typical ratio of hydrodynamic  to  magnetic torque. If we write the resistance tensor in the body frame, $\bar{\D}$, as composed of 4 sub-tensors 
\begin{align}
\bar{\D}=
\begin{bmatrix}
\bar{\D}\mathbf{a}&\bar{\D}\mathbf{b}\\
\bar{\D}\mathbf{b}^T&\bar{\D}\mathbf{c}
\end{bmatrix},
\end{align}
then the final dimensionless   equations to solve  are  given by the system 
\begin{subeqnarray}\label{eq:Newlast_equation1}
&&\slabel{linearsys1}\bar{\D}\mathbf{a}\bar{\U}+\bar{\D}\mathbf{b}\bar{\bOmega}=\0,\\
&&\slabel{linearsys2}{\rm Ma}(\bar{\D}\mathbf{b}^T\bar{\U}+\bar{\D}\mathbf{c}\bar{\bOmega})+\e_y\times\bar{\B}=\0,\\
&&[\bar{\B}]_{lab}=[\cos\bar{t}, \sin\bar{t}, 0]^T,\\
&&\slabel{linearsys3}\frac{d\e_x}{dt}=\bar{\bOmega}\times\e_x,\,
\frac{d\e_y}{dt}=\bar{\bOmega}\times\e_y,\,
\frac{d\e_z}{dt}=\bar{\bOmega}\times\e_z.
\end{subeqnarray}

\begin{figure}[t]
\centering
\includegraphics[width=6in]{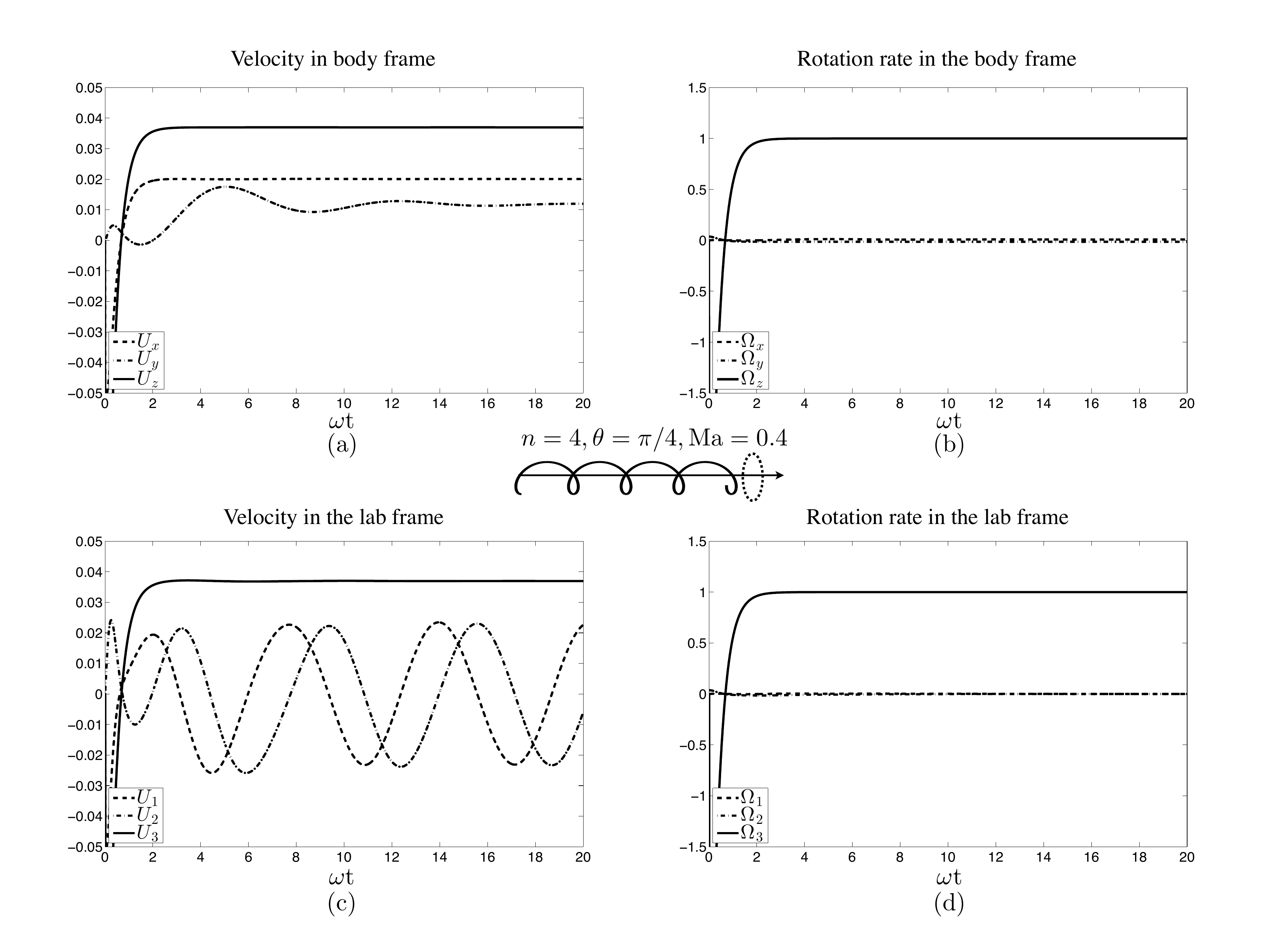}
\caption{
Example of a straight swimming artificial bacteria flagellum. 
Velocity (left) and rotation rate (right) in the body frame (top) and laboratory frame (bottom) with  $n=4$ wavelengths, a helix angle of $\theta=\pi/4$, and $\rm Ma=0.4$. In the body frame, both the velocity and rotation rates reach steady values, while in the laboratory frame the velocities in the $\e_1$ and $\e_2$ directions oscillate around a zero mean. The wobbling angle in this case is $\beta \approx 0.9\degree$ which is almost zero and thus the helix essentially swims  in a  straight line. The four panels display the variation with the dimensionless time of: 
(a) velocity in the body frame;  
(b) rotation rate in the body frame; 
(c) velocity in the laboratory frame;
(d) rotation rate in the laboratory frame. \label{fig:t-v1}}
\end{figure}  
%%%%%%%%%%%%%%%%%%%%%%%%%
\section{Numerical Results}

To address wobbling we first turn to numerical simulations of the system in \eqref{eq:Newlast_equation1}. 
To be relevant to the experiments in Ref.~\cite{NelsonNonideal}, we fix the number of wavelengths, $n$, to be 3 or 4 and we pick  $\gamma=2.3\times10^{-3}$. We vary the helix geometry by addressing four different helix angles, namely
 (${\pi}/{10},{\pi}/{6}, {\pi}/{4},{\pi}/{3}$), and we let  the Mason number, $\rm Ma$,  range from 0.001 to 0.1. {\color{black}When $t=0$, velocity and rotation rate are set to be zero, and the body frame is aligned with the lab frame. In fact, no matter what the initial condition is, as the time goes larger, the solution tends to be the unique periodic state.} The system is solved using a {\color{black}partial} Crank-Nicolson method where, at each time step, the rotation rate is obtained from the linear system, Eqs.~(\ref{linearsys1})-(\ref{linearsys2}), with the information  from the location of the body frame from the previous step. {\color{black}The method is  partial as the rotation rate is explicit in \eqref{linearsys3}.}

\begin{figure}[t]
\centering
\includegraphics[width=6in]{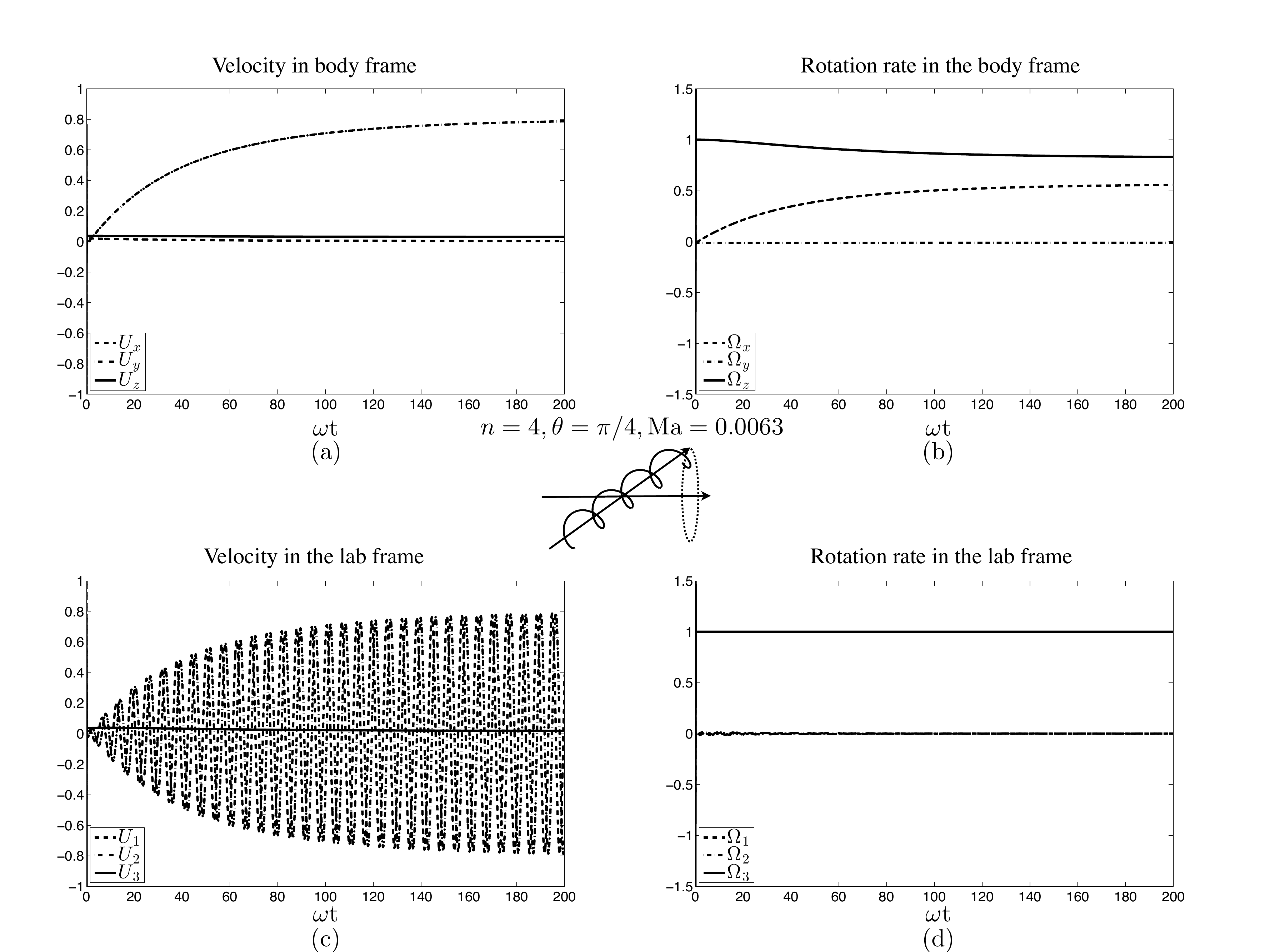}
\caption{Same as Fig.~\ref{fig:t-v1} but with   a Mason number decreased to $\rm Ma=0.0063$. In this case significant wobbling is obtained with  $\beta \approx 35\degree$. Compared to Fig.~\ref{fig:t-v1} the mean velocity in the forward direction has decreased while the  velocity amplitudes in other two directions have  increased.  In addition, the rotation rate in the  $\e_x$ direction is no longer   zero.\label{fig:t-v2}}
\end{figure}

Our numerical simulations demonstrate the experimentally-observed  transition from wobbling at low Mason number to directional swimming at high Mason number. To illustrate this transition we plot in  Figs.~\ref{fig:t-v1} and~\ref{fig:t-v2} the dynamics, both in the body frame (top) and the laboratory frame (bottom) of two  helices displaying qualitatively different behaviors. In Fig.~\ref{fig:t-v1} we show the velocity (left) and rotation rate (right) of a helix with $n=4$ wavelengths and a helix angle of $\theta=\pi/4$ at a Mason number of $\rm Ma=0.4$. The helix is seen to swim straight  (small wobbling angle $\beta\approx 0.9\degree$). In contrast we show in Fig.~\ref{fig:t-v2} the   kinematics for the same helix   at a smaller Mason number of $\rm Ma=0.0063$. In that case, as can clearly be seen in Fig.~\ref{fig:t-v2}c, the components of the helix velocity in the  plane perpendicular to the average swimming direction are time-periodic and of amplitude large compared to the average swimming speed along the third direction. This is an indication of wobbling with a large angle (here, $\beta\approx 35\degree$). Wobbling can also be seen by comparing the values of the rotation rates in the body frame in Fig.~\ref{fig:t-v1}b and Fig.~\ref{fig:t-v2}b. When no wobbling occurs and the helix is swimming straight, the body-frame rotation rate includes only a component in the direction of the helix axis ($z$ direction). In contrast, for a  helix with significant wobbling, a component of the rotation rate perpendicular to the direction of the axis helix exists ($x$ direction), of the same order as the axial rotation rate.

\begin{figure}[t]
\centering
\includegraphics[width=6.5in]{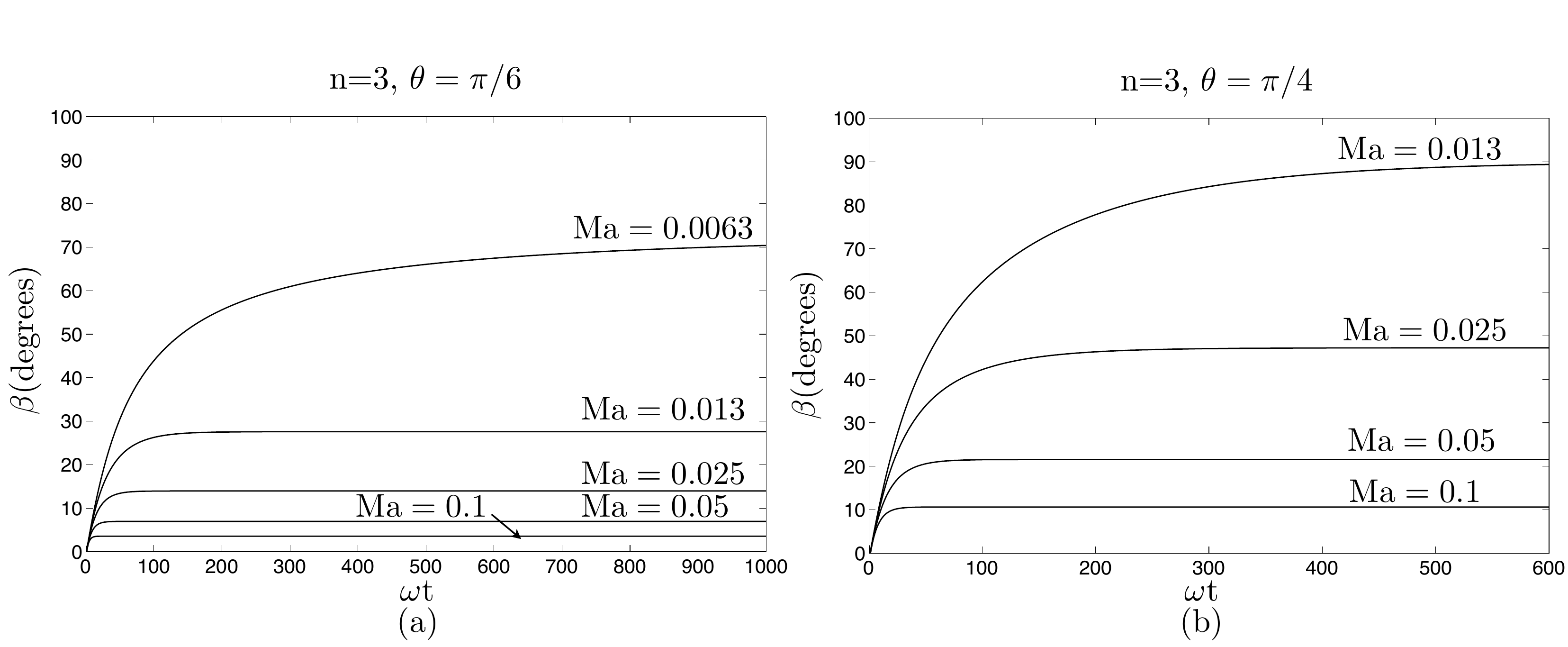}
\caption{Dependence of the wobbling angle $\beta$  (in degrees) on the dimensionless time, $\omega t$, for a helix with $n=3$ wavelengths and an  angle of $\theta=\pi/6$ (a) and $\theta=\pi/4$ (b).\label{fig:t-beta}}
\end{figure}

\begin{figure}[t]
\centering
\includegraphics[width=6.5in]{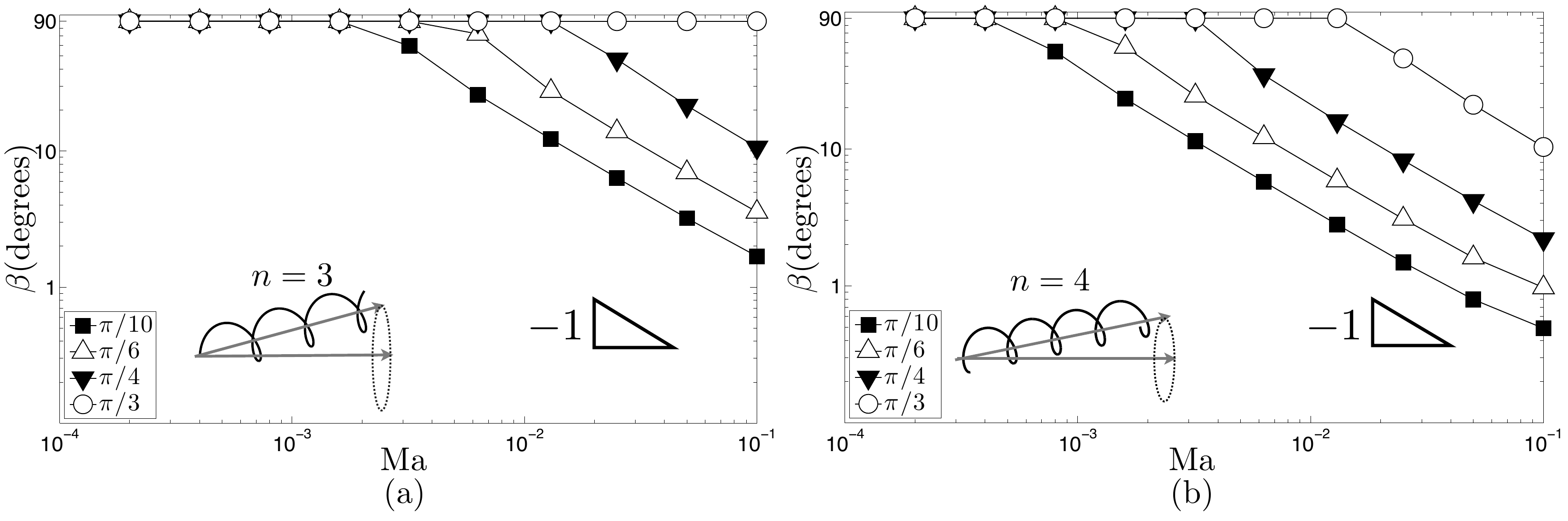}
\caption{Dependence of the long-time wobbling angle $\beta$ (in degrees) on the Mason number, $\rm Ma$, for four different helix angles  (${\pi}/{10},{\pi}/{6}, {\pi}/{4},{\pi}/{3}$) and a number of wavelengths $n=3$ (left) and $n=4$ (right). Results are plotted on a log-log scale demonstrating a $\beta\sim \rm Ma^{-1}$ relationship in the wobbling-to-swimming transition.
\label{fig:beta-ma}}
\end{figure}

To further quantify wobbling, we perform simulations where we measure the time-variation of the wobbling angle. The results for $n=3$ are illustrated in Fig.~\ref{fig:t-beta} for two values of the  helix angles. For all values of the Mason number, we observe convergence of the wobbling angle to a steady value. The dependence of that long-time, steady value on the Mason number is shown in Fig.~\ref{fig:beta-ma} for $n=3$ (left) and $n=4$ (right) and for four values of the helix angle. For every helix, the wobbling angle is 90$\degree$  for {\color{black}low} $\rm Ma$ number while it decreases to zero as $\beta\sim \rm Ma^{-1}$ for large values of the Mason number. This dependence  with $\rm Ma$ is reminiscent of the inverse frequency behavior seen experimentally in Fig.~\ref{fig:experimental} \cite{NelsonNonideal}. For a fixed Mason number, the wobbling-to-swimming transition is also affected by the helix geometry. Specifically, directed swimming happens sooner for helices with larger number of wavelengths ($n$) and smaller helix angles  ($\theta$).

%%%%%%%%%%%%%%%%%%%%%%%%%
\section{Asymptotic Analysis}

Our numerical computations demonstrate the wobbling-to-swimming transition. We saw in particular in the transition region an inverse linear relationship between wobbling angle and Mason number. In this section we present an analytical description of the helix dynamics and recover the $\beta \sim \rm Ma^{-1}$ scaling.  In order to proceed we take  advantage of the fact that if the helix angle is zero, $\theta=0$, the helix turns into a rod which does not swim but for which the steady state dynamics can be solved exactly. We therefore examine the dynamics analytically in the small-$\theta$ limit.

We consider the dimensionless dynamical system given by \eqref{eq:Newlast_equation1} and drop the bars for notation convenience. We pick the number of wavelengths, $n$, to be an integer in order to simplify some of the algebra (although our procedure remains valid for non-integer number of wavelengths). First off, in order to facilitate the expansion, we write  \eqref{eq:Newlast_equation1} component by component as 
\begin{subeqnarray}\label{eq:Newlast_equation2}
\slabel{eq:Newlinearsys1} \left(Da_{ij}U_{j}+Db_{ij}\Omega_{j}\right)\e_i & = & \0,\\
\slabel{eq:Newlinearsys2}{\rm Ma}\left(Db_{ji}U_j+Dc_{ij}\Omega_{j}\right)\e_i+\epsilon_{mnl}e_{yn}B_{l}\e_m&=&\0,\\
\slabel{eq:Newode} \frac{d\e_i}{dt}=\bOmega\times\e_i=\epsilon_{jkp}\Omega_ke_{ip}\e_j&=&\epsilon_{ijk}\Omega_k\e_j.
\end{subeqnarray}
In \eqref{eq:Newlast_equation2},  in order to differentiate between base vectors in the laboratory vs.~body frame we use the following convention:   vectors with  subscripts 
$(m, n, l)$ are in the laboratory frame (therefore $1,2, 3$)
while those with with subscripts 
$(i, j, k, p)$  are  in  the body frame (therefore $x,y,z$).   
As a consequence,  the terms $e_{yn}$ in  \eqref{eq:Newlinearsys2} and $e_{ip}$ in  \eqref{eq:Newode} are different: the first one refers to the components in the laboratory frame of the vector ${\bf e}_y$ while the latter refers to the $p$th components of ${\bf e}_i$ in the body frame, i.e.~$\delta_{ip}$.  In the body frame we  write $\U=U_x\e_x+U_y\e_y+U_z\e_z$ and $\bOmega=\Omega_x\e_x+\Omega_y\e_y+\Omega_z\e_z$, and similarly for each {\color{black}component} of the tensor  $\D$. The advantage of using the body frame is that, in it, the components of $\D$  are all constant. The components of  body frame vector, $\e_i$,  in the laboratory frame are written as $[\e_i]_{lab}=[e_{i1}, e_{i2}, e_{i3}]^T$, for any $i$ is in $(x, y, z)$.  

Aiming at solving \eqref{eq:Newlast_equation2}  order by order, we expand  all variables in powers of $\theta$ as
\begin{eqnarray}\label{eq:expansion}
\{ U_i, \Omega_i, D_{ij}, [\e_i]_{lab}, \xi_{\parallel}\} &=& \{U_i\upl, \Omega_i\upl, D_{ij}\upl, [\e_i]_{lab}\upl, \xi\upl_\parallel\}\notag\\ \notag
&&+ \theta \{U_i\upf, \Omega_i\upf, D_{ij}\upf, [\e_i]_{lab}\upf, \xi\upf_\parallel\} \\
&& + \dots,
\end{eqnarray}
for any $i,j$ in $(x, y, z)$. 
In the body frame, the coefficients of the tensor $\D$, written under matrix form as $\D_{body}$, are  given  in  Appendix \ref{app}.  They involve the helix geometry and the viscous resistance coefficient, $\xi_\parallel$.  The expansion for that coefficient is
 \begin{eqnarray}\label{eq:expansion_xi}
\notag \xi_\parallel&=&\frac{2\pi}{\ln({2\cos\theta}/{\gamma})-{1}/{2}}=2\pi\left[\ln\frac{2}{\gamma}-\frac{1}{2}-\frac{\theta^2}{2}+o(\theta^2)\right]^{-1} \\
&=&\frac{2\pi}{\ln({2}/{\gamma})-{1}/{2}}+\frac{\pi\theta^2}{\left[\ln({2}/{\gamma})-{1}/{2}\right]^2}+o(\theta^2).
\end{eqnarray}
We therefore obtain $\xi\upl_\parallel={2\pi}/[{\ln({2}/{\gamma})-{1}/{2}}]$,  $\xi\upf_\parallel=0$, and $\xi\ups_\parallel= {\pi}/{\left[\ln({2}/{\gamma})-{1}/{2}\right]^2}$. Using the expressions in Appendix \ref{app}, it follows that 
\begin{gather}\label{eq:D0}
[\D]_{body}\upl=\xi\upl_\parallel
\begin{bmatrix}       
-2n&0&0&0&-n^2&0\\
0&-2n&0&n^2&0&0\\
0&0&-n&0&0&0\\
0&n^2&0&-\frac{2n^3}{3}&0&0\\
-n^2&0&0&0&-\frac{2n^3}{3}&0\\
0&0&0&0&0&\frac{4\pi\gamma^2}{\xi\upl_\parallel} n
\end{bmatrix} 
\end{gather}
and 
\begin{gather}\label{eq:D1}
[\D]_{body}\upf=\xi\upl_\parallel
\begin{bmatrix}
0 & 0&0&0&0&0\\
0&0&0&0&0&0\\
0&0&0&0&\frac{n}{2\pi}&0\\
0&0&0&0&0&0\\
0&0&\frac{n}{2\pi}&0&0&-\frac{n}{2\pi^2}\\
0&0&0&0&-\frac{n}{2\pi^2}&0
\end{bmatrix}.
\end{gather}
%%%%
\subsection{Zeroth order solution} 
At  zeroth order, the helix becomes a rigid rod. In that case, \eqref{eq:Newlast_equation2} becomes
\begin{subeqnarray}\label{eq:Newlast_equation2_zeroth}
 \left(Da_{ij}\upl U_{j}\upl+Db_{ij}\upl\Omega_{j}\upl\right)\e_i\upl&=&\0,\\
{\rm Ma}\left(Db_{ji}\upl U_j\upl+Dc_{ij}\upl\Omega_{j}\upl\right)\e_i\upl+\epsilon_{mnl}e_{yn}\upl B_{l}\e_m&=&\0,\\
 \frac{d\e_i\upl}{dt}&=&\epsilon_{ijk}\Omega_k\upl\e_j\upl.
\end{subeqnarray}
The obvious steady solution to \eqref{eq:Newlast_equation2_zeroth} is then given by
\begin{subeqnarray}\label{eq:Newsolution_zeroth}
U_x\upl& = &U_y\upl = U_z\upl = 0,\\
\Omega_x\upl& = &\Omega_y\upl =0,\quad \Omega_z\upl=1,\\
{}[\e_{x}]_{lab}\upl& = &[\sin (t+\psi_0),-\cos (t+\psi_0),0]^T,\\
{}[\e_{y}]_{lab}\upl& = &[\cos (t+\psi_0),\sin(t+\psi_0),0]^T,\\
{}[\e_{z}]_{lab}\upl& = &[0,0,1]^T,
\end{subeqnarray}
where $\psi_0$, satisfying 
\begin{equation}
\sin\psi_0=4\pi\gamma^2n\rm Ma,
\end{equation}
is the phase delay between the applied field and the rotation of the rod. At leading order, the rod simply is aligned with, and rotates around, the axis perpendicular to the plane of the applied field with no wobbling.

%%%%
\subsection{First order solution}

At order $O(\theta)$, \eqref{eq:Newlast_equation2} become 
\begin{subeqnarray}\label{eq:Newlast_equation2_first}
\0 &=& \left(Da_{ij}\upl U_{j}\upf+Db_{ij}\upl\Omega_{j}\upf+Da_{ij}\upf U_{j}\upl+Db_{ij}\upf\Omega_{j}\upl\right)\e_i\upl\nonumber  \\
&& \slabel{eq:Newlinearsys1_first}+\left(Da_{ij}\upl U_{j}\upl+Db_{ij}\upl\Omega_{j}\upl\right)\e_i\upf,\\
\0 &=&\nonumber {\rm Ma}\left(Db_{ji}\upl U_j\upf+Dc_{ij}\upl\Omega_{j}\upf+Db_{ji}\upf U_j\upl+Dc_{ij}\upf\Omega_{j}\upl\right)\e_i\upl  \\
&&\slabel{eq:Newlinearsys2_first}+{\rm Ma}\left(Db_{ji}\upl U_j\upl+Dc_{ij}\upl\Omega_{j}\upl\right)\e_i\upf+\epsilon_{mnl}e_{yn}\upf B_{l}\e_m,\\
\slabel{eq:Newode_first} \frac{d\e_i\upf}{dt}&=&\epsilon_{ijk}\left(\Omega_k\upl\e_j\upf+\Omega_k\upf\e_j\upl\right).
\end{subeqnarray}

The system of equation for the first-order unknowns in  \eqref{eq:Newlast_equation2_first} is linear and can thus be solved exactly. Using \eqref{eq:Newlinearsys1_first}, the number of unknowns can be decreased by three
\begin{equation}\label{eq:Newvelocity_first}
U_x\upf = -\frac{n}{2}\Omega_y\upf, \quad U_y\upf = \frac{n}{2}\Omega_x\upf, \quad U_z\upf = 0.
\end{equation}
Then using \eqref{eq:Newlinearsys2_first} and combining it with \eqref{eq:Newvelocity_first}, the rotational components  can be obtained as the function of the components of body frame vectors expressed  in the laboratory frame as
\begin{subeqnarray}\label{eq:Newomega_first}
\Omega_{x}\upf&=&-\frac{6}{\xi\upl_\parallel n^3{\rm Ma}}\left[e_{y3}\upf\cos\psi_0-e_{z1}\upf\sin\psi_0\sin(t+\psi_0)+e_{z2}\upf\sin\psi_0\cos(t+\psi_0)\right],\\
\Omega_{y}\upf&=&-\frac{3}{\pi^2n^2}+\frac{6\sin\psi_0}{\xi\upl_\parallel n^3{\rm Ma}}\left[e_{y3}\upf+e_{z1}\upf\cos(t+\psi_0)+e_{z2}\upf\sin(t+\psi_0)\right],\\
\Omega_{z}\upf&=&-e_{z3}\upf+\frac{1}{\sin\psi_0}(e_{y2}\upf\cos t-e_{y1}\upf\sin t).
\end{subeqnarray}
Finally substituting \eqref{eq:Newomega_first} into \eqref{eq:Newode_first}, we obtain the full order-one solution as
\begin{subeqnarray}\label{eq:Newsolution_first}
\slabel{Omega1}\Omega_x\upf& = &\frac{18\cos\psi_0}{\pi^2\xi\upl_\parallel n^5{\rm Ma}\left(1+\frac{24\pi\gamma^2}{\xi\upl_\parallel n^2}\right)},
\quad \Omega_y\upf = -\frac{3}{\pi^2n^2},
\quad \Omega_z\upf = 0,\\
{}[\e_x]_{lab}\upf & = & \left[0, 0, \frac{18\cos\psi_0}{\pi^2\xi\upl_\parallel n^5{\rm Ma}
\left(1+\frac{24\pi\gamma^2}{\xi\upl_\parallel n^2}\right)}\right]^T,\\
{}[\e_y]_{lab}\upf & = & \left[ 0,0,-\frac{3}{\pi^2n^2} \right]^T,\\
{}[\e_z]_{lab}\upf & = & -\sqrt{[\Omega_x\upf]^2+[\Omega_y\upf]^2}\left[ \cos(t+\psi_0-\psi_1),
\sin(t+\psi_0-\psi_1),0\right]^T,\slabel{ez1}
\end{subeqnarray}
with $\tan\psi_1=\Omega_x\upf/\Omega_y\upf$.

%%%%%%%%%%%%%%%%
\subsection{Wobbling angle}
With the knowledge of both the zeroth and first-order solution we can now calculate our prediction for the wobbling angle, $\beta$. It is defined as $\sin\beta=\sqrt{e_{z1}^2+e_{z2}^2}$. Since the zeroth-order solution shows no wobbling, we have $\beta =  O(\theta)$ and thus can use the approximation $\sin\beta\approx\beta$. Given  \eqref{ez1}
we therefore obtain  $\beta \approx \theta \left({\left[\Omega_x\upf\right]^2+\left[\Omega_y\upf\right]^2}\right)^{1/2}$ which, using \eqref{Omega1}, becomes
\begin{equation}
\beta\approx \frac{3 \theta }{\pi^2 n^2}
\left[
\left(
\frac{6\cos\psi_0}{\xi\upl_\parallel n^3{\rm Ma}\left(1+\frac{24\pi\gamma^2}{\xi\upl_\parallel n^2}\right)}
\right)^2
+ 1
\right]^{1/2}.\label{finalmodel}
\end{equation}

Our model, \eqref{finalmodel},  predicts that the wobbling angle decreases with the $\rm Ma$ number, increases with the helix angle, and decreases for large number of wavelengths $n$, which are that the three main observations from our computational results. In addition, for  low values of  $\rm Ma$ we get from  \eqref{finalmodel} the approximate angle
\begin{equation}\label{eq:wobbling2}
\beta\approx \frac{18\theta\cos\psi_0}{\pi^2\xi\upl_\parallel n^5{\rm Ma}\left(1+\frac{24\pi\gamma^2}{\xi\upl_\parallel n^2}\right)}\cdot
\end{equation}
This can be further simplified by noting that $\cos\psi_0 = \sqrt{1-16\pi^2\gamma^4n^2{\rm Ma}^2}\approx 1$ at low $\rm Ma$. In addition, 
$24 \pi \gamma^2/ \xi\upl_\parallel n^2 \approx 12  \gamma^2 [{\ln({2}/{\gamma})-{1}/{2}}] / n^2 \ll 1$ for $\gamma \ll 1$. 
Our approximation, \eqref{eq:wobbling2}, can therefore be further simplified as
\begin{equation}
\beta  \approx \frac{18\theta}{\pi^2\xi\upl_\parallel n^5\rm Ma}
\approx \frac{9[{\ln({2}/{\gamma})-{1}/{2}}]}{\pi^3}\frac{\theta}{ n^5\rm Ma}\cdot
\label{betafinal}
\end{equation}
The theoretical approach allows therefore to recover the $\beta\sim \rm Ma^{-1}$ scaling observed experimentally and numerically  in the wobbling-to-swimming transition. 

%%%%%%%%%%%%%%%%%%
\subsection{Time-averaged swimming velocity}
Using our model, we can {\color{black}go} beyond the prediction for the wobbling angle and  calculate the  time-averaged swimming velocity of the helix at leading order.  The swimming speed is written as $\U = U_i\e_i$ in the body frame, which can be expended as
\begin{align}
\U= U_i\upl\e_i\upl+\theta\left(U_i\upl\e_i\upf+U_i\upf\e_i\upl\right)+\theta^2\left(U_i\upl\e_i\ups+U_i\upf\e_i\upf+U_i\ups\e_i\upl\right)+...
\end{align}
where  $i$ is in  $(x, y, z)$. The forward velocity of interest is the component  $U_3$ along the  direction perpendicular to the applied magnetic field. 
With the information from 
Eqs.~(\ref{eq:Newsolution_zeroth}), (\ref{eq:Newvelocity_first}), and (\ref{eq:Newsolution_first}), we get  the velocity in the laboratory frame.
\begin{subeqnarray}\label{eq:Newvelocity_lab}
U_1& = & -\frac{\theta n}{2}\sqrt{\left[\Omega_x\upf\right]^2+\left[\Omega_y\upf\right]^2}\sin(t+\psi_0)+o(\theta),\\
U_2& = & \frac{\theta n}{2}\sqrt{\left[\Omega_x\upf\right]^2+\left[\Omega_y\upf\right]^2}\cos(t+\psi_0)+o(\theta),\\
U_3& = & o(\theta),
\end{subeqnarray}
and therefore we have to go to the next order in $\theta$ to obtain the leading-order behavior for  $U_3$. At order $O(\theta^2)$ we have
\begin{equation}
\U\ups = U_i\upl\e_i\ups+U_i\upf\e_i\upf+U_i\ups\e_i\upl,
\end{equation}
and given that we know that $U_i\upl=U_z\upf=e_{x3}\upl=e_{y3}\upl=0$, and $e_{z3}\upl=1$, we obtain
\begin{equation}\label{eq:Newvelocity_second1}
U_3\ups=U_x\upf e_{x3}\upf+U_y\upf e_{y3}\upf+U_z\ups=U_z\ups,\\
\end{equation}
which means we only need to  solve for $U_z\ups$. At second order, \eqref{eq:Newlinearsys1} becomes 
\begin{eqnarray}
\nonumber\left(Da_{ij}\upl U_{j}\ups+Db_{ij}\upl\Omega_{j}\ups+Da_{ij}\upf U_{j}\upf+Db_{ij}\upf\Omega_{j}\upf+Da_{ij}\ups U_{j}\upl+Db_{ij}\ups\Omega_{j}\upl\right)\e_i\upl&\\
\nonumber+\left(Da_{ij}\upl U_{j}\upf+Db_{ij}\upl\Omega_{j}\upf+Da_{ij}\upf U_{j}\upl+Db_{ij}\upf\Omega_{j}\upl\right)\e_i\upf&\\ 
+\left(Da_{ij}\upl U_{j}\upl+Db_{ij}\upl\Omega_{j}\upl\right)\e_i\ups&=\0.
\label{eq:Newlinearsys1_second}\end{eqnarray}

Combining the solutions in Eqs.~(\ref{eq:Newsolution_zeroth}), (\ref{eq:Newvelocity_first}) and (\ref{eq:Newsolution_first}), we obtain the simplifications 
\begin{subeqnarray}
\left(Da_{ij}\upl U_{j}\upl+Db_{ij}\upl\Omega_{j}\upl\right)\e_i\ups&=&\0,\\ \left(Da_{ij}\upl U_{j}\upf+Db_{ij}\upl\Omega_{j}\upf\right)\e_i\upf&=&\0,\\\left(Da_{ij}\upf U_{j}\upl+Db_{ij}\upf\Omega_{j}\upl\right)\e_i\upf&=&\0.
\end{subeqnarray}
As we have $Da_{ij}\upf=U_i\upl=0$, the corresponding terms  cancel out, and \eqref{eq:Newlinearsys1_second}
 simplifies to
\begin{equation}\label{eq:Newlinearsys1_second_final}
\left(Da_{ij}\upl U_{j}\ups+Db_{ij}\upl\Omega_{j}\ups+Db_{ij}\upf\Omega_{j}\upf+Db_{ij}\ups\Omega_{j}\upl\right)\e_i\upl=\0.
\end{equation}
The second order expansion of $[\Db]_{body}$ is
\begin{gather}\label{eq:Db2}
[\Db]_{body}\ups=\xi\upl_\parallel
\begin{bmatrix}
-\frac{3}{8\pi}n&\frac{3}{4}n^2&0\\
-\frac{3}{4}n^2&-\frac{1}{8\pi}n&0\\
0&0&\frac{n}{2\pi}
\end{bmatrix}
+\xi\ups_\parallel
\begin{bmatrix}
0&-n^2&0\\
n^2&0&0\\
0&0&0
\end{bmatrix}.
\end{gather}
Substituting \eqref{eq:Db2} into \eqref{eq:Newlinearsys1_second_final}, we obtain
\begin{equation}
-nU_z\ups+\frac{n}{2\pi}\Omega_y\upf+\frac{n}{2\pi}=0.
\end{equation}
With \eqref{eq:Newvelocity_second1}, this finally leads to the leading-order expression for the time-averaged swimming speed in the laboratory frame as
\begin{equation}
U_3\ups=\frac{1}{2\pi}\left(1-\frac{3}{\pi^2n^2}\right)\cdot
\end{equation}
Note that we have $U_3=O(\theta^2)$, while both $U_1$ and $U_2$ are $O(\theta)$, and thus for a small helix angle the forward swimming velocity is much smaller then the velocities perpendicular to the average swimming direction.

%%%%%%%%%%%%%%%%%%%%%%%%%
\section{Discussion}

Motivated by {\color{black}experiments} on artificial bacterial flagella driven by external magnetic fields we address theoretically in this paper the dynamics of rigid helices under time-periodic torques. Using numerical computations we obtain, in agreement with experimental results, that driven helices do swim in a directed fashion at high Mason number but wobble around their mean swimming direction for lower values of the Mason number. During the  wobbling-to-swimming transition, the wobbling angle varies as the inverse of Mason number. The shape of the helix also affects this transition and helices with larger number of wavelengths or smaller helix angle  are seen to swim {\color{black}more efficiently}.  We then use an asymptotic analysis of the helix dynamics in the near-rod geometric limit to derive analytically an estimate for the wobbling angle. Our prediction, \eqref{betafinal}, shows the same inverse $\rm Ma$ dependence as in our computations and experimental work, and quantifies the strong influence of the  helix geometry on the swimming behavior. 

We hope our results will help  guide the future design of  artificial bacterial flagella. Two  factors important for the practical operation of micro-swimmers are that they remain stable while being efficient energetically.  Energy efficiency is   bound to play an important role since any external source of power not converted to useful propulsive work will be dissipated in the surrounding fluid, leading to heating which is possibly problematic in biological environments.  As is well known, a rotating rigid helix  is most efficient when its helix angle, $\theta$, is around 40 degrees \cite{lauga09}. Stability was addressed specifically in our paper, and we now understand the dynamic and energetic factors impacting it. 
From \eqref{betafinal}, we have $\beta \sim \theta/n^5 {\rm Ma}$ and 
 we see that, with the value of $\theta$ fixed,  stability of swimming (i.e non-wobbling) will be obtained for large values of $n$ and $\rm Ma$. Recalling that ${\rm Ma}={\mu\omega\Lambda^3}/{ HB_0}$, we get a wobbling angle scaling as $\beta \sim \theta  HB_0 / {\mu\omega n^5 \Lambda^3}$.  {\color{black}Perhaps counter-intuitively}, wobbling is avoided when small  magnetic field and magnetic moments are used. Low frequencies should also be avoided. In addition, given that the total helix length is $L\sim n\Lambda$, we see that long helices, with many wavelengths, are preferable. 
 
 Of course the work considered here only addressed the case of a single artificial bacterial flagellum, and it could be that swimmers composed or more than one helices would be more stable. For example, two identical parallel and counter-rotating helices are stable in the plane containing the two helix axis, which  would therefore constraint  potential wobbling to the plane perpendicular to it. A combination of such helix pair with its mirror image would be  stable and always swim along a straight line,  but such elaborate geometry would probably require infinite ingenuity to be implemented in practice. Decreasing length scales even further to design nanometer-scale swimmers will further complicate the dynamics by introducing thermal fluctuations.    The hunt for the ultimate stable and efficient artificial low-Reynolds swimmer is still open.

%%%%%%%%%%%%%%%%%%%%%%%%%
\begin{acknowledgements}
We thank  B. Nelson and his research group  at ETH Zurich for stimulating discussions which initiated our interest in the topic of artificial bacteria flagella. Funding by the National Science Foundation (grant CBET-0746285) is gratefully acknowledged.
\end{acknowledgements}

%%%%%%%%%%%%%%%%%%%%%%%%%
\appendix
\section{mobilities}\label{app}
All 21 terms of the symmetric viscous resistance matrix, $[\bar{\D}]_{body}=\bar{\xi}_{\parallel}\mathbf{M}$, are given below; the remaining 15 terms can be found using $\mathbf{M}=\mathbf{M}^T.$  We use the notation $\alpha=\cos\theta$, $2\pi\eta=\sin\theta$, $\phi=2\pi n$, and $n$ is any positive number.

\noindent 
\begin{subeqnarray}
M_{11}&=&-\frac{\pi}{2}\eta^2\sin 2\phi-2(1-\pi^2\eta^2)n\\
M_{12}&=&-\frac{\pi}{2}\eta^2(1-\cos2\phi)\\
M_{13}&=&-\alpha\eta(1-\cos\phi)\\
M_{14}&=&-\pi\alpha\eta^2n\left(1+\frac{\cos\phi}{2}\right)+\frac{3}{8}\alpha\eta^2\sin2\phi\\
M_{15}&=&\pi\alpha\eta^2n\left(n\pi-\frac{\sin\phi}{2}\right)+\frac{3}{8}\alpha\eta^2(1-\cos2\phi)-\alpha n^2\\
M_{16}&=&\frac{\eta}{\pi}(1-2\pi^2\eta^2)(1-\cos\phi)\\
M_{22}&=&\frac{\pi}{2}\eta^2\sin 2\phi-2(1-\pi^2\eta^2)n\\
M_{23}&=&\alpha\eta\sin\phi\\
M_{24}&=&-\pi\alpha\eta^2n\left(n\pi+\frac{\sin\phi}{2}\right)+\frac{3}{8}\alpha\eta^2(1-\cos2\phi)+\alpha n^2\\
M_{25}&=&-\pi\alpha\eta^2n\left(1-\frac{\cos\phi}{2}\right)-\frac{3}{8}\alpha\eta^2\sin2\phi\\
M_{26}&=&-\frac{\eta}{\pi}(1-2\pi^2\eta^2)\sin\phi\\
M_{33}&=&(\alpha^2-2)n\\
M_{34}&=&-\alpha^2\eta n\sin\phi-4\pi\eta^3(1-\cos\phi)\\
M_{35}&=&\alpha^2\eta n\cos\phi+4\pi\eta^3\sin\phi\\
M_{36}&=&2\pi\alpha\eta^2 n\\
M_{44}&=&\frac{3}{4}\alpha^2\eta^2n\cos\phi+\left(\frac{\eta^2}{4\pi}+\frac{\pi}{2}\alpha^2\eta^2n^2-\frac{5}{16\pi}\alpha^2\eta^2\right)\sin2\phi-\left(1-\frac{\alpha^2}{2}\right)\eta^2n\\
&&-\frac{2}{3}\alpha^2\left(1-\pi^2\eta^2\right)n^3+\frac{8\pi^3\gamma^2}{\bar{\xi}_{\parallel}}\eta^2\left(n-\frac{\sin2\phi}{4\pi}\right)\\
M_{45}&=&\frac{1}{4}\alpha^2\eta^2n\sin\phi+\left(\frac{\eta^2}{4\pi}+\frac{\pi}{2}\alpha^2\eta^2n^2-\frac{3}{16\pi}\alpha^2\eta^2\right)\left(1-\cos\phi\right)+\frac{\pi}{2}\alpha^2\eta^2n^2\\
&&-\frac{2\pi^2\gamma^2}{\bar{\xi}_{\parallel}}\eta^2\left(1-\cos2\phi\right)\\
M_{46}&=&\frac{\alpha\eta n }{\pi}\left(1-2\pi^2\eta^2\right)\sin\phi-\frac{\alpha^3\eta}{2\pi^2}\left(1-\cos\phi\right)-\frac{4\pi\gamma^2}{\bar{\xi}_{\parallel}}\alpha\eta\left(1-\cos\phi\right)\\
M_{55}&=&-\frac{3}{4}\alpha^2\eta^2n\cos\phi-\left(\frac{\eta^2}{4\pi}+\frac{\pi}{2}\alpha^2\eta^2n^2-\frac{5}{16\pi}\alpha^2\eta^2\right)\sin2\phi-\left(1-\frac{\alpha^2}{2}\right)\eta^2n\\
&&-\frac{2}{3}\alpha^2(1-\pi^2\eta^2)n^3+\frac{8\pi^3\gamma^2}{\bar{\xi}_{\parallel}}\eta^2(n+\frac{\sin2\phi}{4\pi})\\
M_{56}&=&-\frac{\alpha\eta n }{\pi}(1-2\pi^2\eta^2)\cos\phi+\frac{\alpha^3\eta}{2\pi^2}\sin\phi+\frac{4\pi\gamma^2}{\bar{\xi}_{\parallel}}\alpha\eta\sin\phi\\
M_{66}&=&-2\eta^2n(1-2\pi^2\eta^2)+\frac{4\pi\gamma^2}{\bar{\xi}_{\parallel}}\alpha^2n
\end{subeqnarray}

\bibliography{ABF_new}

%merlin.mbs aipnum4-1.bst 2010-07-25 4.21a (PWD, AO, DPC) hacked
%Control: key (0)
%Control: author (8) initials jnrlst
%Control: editor formatted (1) identically to author
%Control: production of article title (0) allowed
%Control: page (1) range
%Control: year (1) truncated
%Control: production of eprint (0) enabled
\begin{thebibliography}{36}%
\makeatletter
\providecommand \@ifxundefined [1]{%
 \@ifx{#1\undefined}
}%
\providecommand \@ifnum [1]{%
 \ifnum #1\expandafter \@firstoftwo
 \else \expandafter \@secondoftwo
 \fi
}%
\providecommand \@ifx [1]{%
 \ifx #1\expandafter \@firstoftwo
 \else \expandafter \@secondoftwo
 \fi
}%
\providecommand \natexlab [1]{#1}%
\providecommand \enquote  [1]{``#1''}%
\providecommand \bibnamefont  [1]{#1}%
\providecommand \bibfnamefont [1]{#1}%
\providecommand \citenamefont [1]{#1}%
\providecommand \href@noop [0]{\@secondoftwo}%
\providecommand \href [0]{\begingroup \@sanitize@url \@href}%
\providecommand \@href[1]{\@@startlink{#1}\@@href}%
\providecommand \@@href[1]{\endgroup#1\@@endlink}%
\providecommand \@sanitize@url [0]{\catcode `\\12\catcode `\$12\catcode
  `\&12\catcode `\#12\catcode `\^12\catcode `\_12\catcode `\%12\relax}%
\providecommand \@@startlink[1]{}%
\providecommand \@@endlink[0]{}%
\providecommand \url  [0]{\begingroup\@sanitize@url \@url }%
\providecommand \@url [1]{\endgroup\@href {#1}{\urlprefix }}%
\providecommand \urlprefix  [0]{URL }%
\providecommand \Eprint [0]{\href }%
\providecommand \doibase [0]{http://dx.doi.org/}%
\providecommand \selectlanguage [0]{\@gobble}%
\providecommand \bibinfo  [0]{\@secondoftwo}%
\providecommand \bibfield  [0]{\@secondoftwo}%
\providecommand \translation [1]{[#1]}%
\providecommand \BibitemOpen [0]{}%
\providecommand \bibitemStop [0]{}%
\providecommand \bibitemNoStop [0]{.\EOS\space}%
\providecommand \EOS [0]{\spacefactor3000\relax}%
\providecommand \BibitemShut  [1]{\csname bibitem#1\endcsname}%
\let\auto@bib@innerbib\@empty
%</preamble>
\bibitem [{\citenamefont {Brennen}\ and\ \citenamefont
  {Winetl}(1977)}]{brennen}%
  \BibitemOpen
  \bibfield  {author} {\bibinfo {author} {\bibfnamefont {C.}~\bibnamefont
  {Brennen}}\ and\ \bibinfo {author} {\bibfnamefont {H.}~\bibnamefont
  {Winetl}},\ }\bibfield  {title} {\enquote {\bibinfo {title} {Fluid mechanics
  of propulsion by cilia and flagella},}\ }\href@noop {} {\bibfield  {journal}
  {\bibinfo  {journal} {Annu. Rev. Fluid Mech.}\ }\textbf {\bibinfo {volume}
  {9}},\ \bibinfo {pages} {339--98} (\bibinfo {year} {1977})}\BibitemShut
  {NoStop}%
\bibitem [{\citenamefont {Pedley}\ and\ \citenamefont
  {Kessler}(1992)}]{pedley92}%
  \BibitemOpen
  \bibfield  {author} {\bibinfo {author} {\bibfnamefont {T.~J.}\ \bibnamefont
  {Pedley}}\ and\ \bibinfo {author} {\bibfnamefont {J.~O.}\ \bibnamefont
  {Kessler}},\ }\bibfield  {title} {\enquote {\bibinfo {title} {Hydrodynamic
  phenomena in suspensions of swimming microorganisms},}\ }\href@noop {}
  {\bibfield  {journal} {\bibinfo  {journal} {Annu. Rev. Fluid Mech.}\ }\textbf
  {\bibinfo {volume} {24}},\ \bibinfo {pages} {313--358} (\bibinfo {year}
  {1992})}\BibitemShut {NoStop}%
\bibitem [{\citenamefont {Nelson}, \citenamefont {Kaliakatsos},\ and\
  \citenamefont {Abbott}(2010)}]{nelson10}%
  \BibitemOpen
  \bibfield  {author} {\bibinfo {author} {\bibfnamefont {B.~J.}\ \bibnamefont
  {Nelson}}, \bibinfo {author} {\bibfnamefont {I.~K.}\ \bibnamefont
  {Kaliakatsos}}, \ and\ \bibinfo {author} {\bibfnamefont {J.~J.}\ \bibnamefont
  {Abbott}},\ }\bibfield  {title} {\enquote {\bibinfo {title} {Microrobots for
  minimally invasive medicine},}\ }\href@noop {} {\bibfield  {journal}
  {\bibinfo  {journal} {Annu. Rev. Biomed. Eng.}\ }\textbf {\bibinfo {volume}
  {12}},\ \bibinfo {pages} {55--85} (\bibinfo {year} {2010})}\BibitemShut
  {NoStop}%
\bibitem [{\citenamefont {Abbott}\ \emph {et~al.}(2009)\citenamefont {Abbott},
  \citenamefont {Peyer}, \citenamefont {Lagomarsino}, \citenamefont {Zhang},
  \citenamefont {Dong}, \citenamefont {Kaliakatsos},\ and\ \citenamefont
  {Nelson}}]{abbot}%
  \BibitemOpen
  \bibfield  {author} {\bibinfo {author} {\bibfnamefont {J.}~\bibnamefont
  {Abbott}}, \bibinfo {author} {\bibfnamefont {K.}~\bibnamefont {Peyer}},
  \bibinfo {author} {\bibfnamefont {M.}~\bibnamefont {Lagomarsino}}, \bibinfo
  {author} {\bibfnamefont {L.}~\bibnamefont {Zhang}}, \bibinfo {author}
  {\bibfnamefont {L.}~\bibnamefont {Dong}}, \bibinfo {author} {\bibfnamefont
  {I.}~\bibnamefont {Kaliakatsos}}, \ and\ \bibinfo {author} {\bibfnamefont
  {B.}~\bibnamefont {Nelson}},\ }\bibfield  {title} {\enquote {\bibinfo {title}
  {How should microrobots swim?}}\ }\href@noop {} {\bibfield  {journal}
  {\bibinfo  {journal} {Int. J. Robot. Res.}\ }\textbf {\bibinfo {volume}
  {28}},\ \bibinfo {pages} {1434} (\bibinfo {year} {2009})}\BibitemShut
  {NoStop}%
\bibitem [{\citenamefont {Kosa}\ \emph {et~al.}(2011)\citenamefont {Kosa},
  \citenamefont {Jakab}, \citenamefont {Szekely},\ and\ \citenamefont
  {Hata}}]{kosa}%
  \BibitemOpen
  \bibfield  {author} {\bibinfo {author} {\bibfnamefont {G.}~\bibnamefont
  {Kosa}}, \bibinfo {author} {\bibfnamefont {P.}~\bibnamefont {Jakab}},
  \bibinfo {author} {\bibfnamefont {G.}~\bibnamefont {Szekely}}, \ and\
  \bibinfo {author} {\bibfnamefont {N.}~\bibnamefont {Hata}},\ }\bibfield
  {title} {\enquote {\bibinfo {title} {Mri driven magnetic microswimmers},}\
  }\href@noop {} {\bibfield  {journal} {\bibinfo  {journal} {Biomed.
  Microdevices}\ }\textbf {\bibinfo {volume} {14}},\ \bibinfo {pages} {165}
  (\bibinfo {year} {2011})}\BibitemShut {NoStop}%
\bibitem [{\citenamefont {Kosa}\ \emph {et~al.}(2008)\citenamefont {Kosa},
  \citenamefont {Jakab}, \citenamefont {Hata}, \citenamefont {Jolesz},
  \citenamefont {Neubach}, \citenamefont {Shoham},\ and\ \citenamefont
  {Menashe}}]{Kosa08}%
  \BibitemOpen
  \bibfield  {author} {\bibinfo {author} {\bibfnamefont {G.}~\bibnamefont
  {Kosa}}, \bibinfo {author} {\bibfnamefont {P.}~\bibnamefont {Jakab}},
  \bibinfo {author} {\bibfnamefont {N.}~\bibnamefont {Hata}}, \bibinfo {author}
  {\bibfnamefont {F.}~\bibnamefont {Jolesz}}, \bibinfo {author} {\bibfnamefont
  {Z.}~\bibnamefont {Neubach}}, \bibinfo {author} {\bibfnamefont
  {M.}~\bibnamefont {Shoham}}, \ and\ \bibinfo {author} {\bibnamefont
  {Menashe}},\ }\bibfield  {title} {\enquote {\bibinfo {title} {Flagellar
  swimming for medical micro robots: Theory, experiments and application},}\
  }in\ \href@noop {} {\emph {\bibinfo {booktitle} {Proceedings of the 2nd
  Biennial IEEE/RAS-EMBS International Conference on Biomedical Robotics and
  Biomechatronics}}}\ (\bibinfo {address} {Scottsdale, AZ, USA},\ \bibinfo
  {year} {2008})\ pp.\ \bibinfo {pages} {258--263}\BibitemShut {NoStop}%
\bibitem [{\citenamefont {Purcell}(1977)}]{purcell}%
  \BibitemOpen
  \bibfield  {author} {\bibinfo {author} {\bibfnamefont {E.}~\bibnamefont
  {Purcell}},\ }\bibfield  {title} {\enquote {\bibinfo {title} {Life at low
  reynolds number},}\ }\href@noop {} {\bibfield  {journal} {\bibinfo  {journal}
  {Am. J. Phys.}\ }\textbf {\bibinfo {volume} {45}},\ \bibinfo {pages} {3--11}
  (\bibinfo {year} {1977})}\BibitemShut {NoStop}%
\bibitem [{\citenamefont {Lauga}\ and\ \citenamefont {Powers}(2009)}]{lauga09}%
  \BibitemOpen
  \bibfield  {author} {\bibinfo {author} {\bibfnamefont {E.}~\bibnamefont
  {Lauga}}\ and\ \bibinfo {author} {\bibfnamefont {T.}~\bibnamefont {Powers}},\
  }\bibfield  {title} {\enquote {\bibinfo {title} {The hydrodynamics of
  swimming microorganisms},}\ }\href@noop {} {\bibfield  {journal} {\bibinfo
  {journal} {Rep. Prog. Phys.}\ }\textbf {\bibinfo {volume} {72}},\ \bibinfo
  {pages} {096601} (\bibinfo {year} {2009})}\BibitemShut {NoStop}%
\bibitem [{\citenamefont {Lauga}(2011)}]{lauga11}%
  \BibitemOpen
  \bibfield  {author} {\bibinfo {author} {\bibfnamefont {E.}~\bibnamefont
  {Lauga}},\ }\bibfield  {title} {\enquote {\bibinfo {title} {Life around the
  scallop theorem},}\ }\href {\doibase 10.1039/C0SM00953A} {\bibfield
  {journal} {\bibinfo  {journal} {Soft Matter}\ }\textbf {\bibinfo {volume}
  {7}},\ \bibinfo {pages} {3060--3065} (\bibinfo {year} {2011})}\BibitemShut
  {NoStop}%
\bibitem [{\citenamefont {Paxton}\ \emph {et~al.}(2004)\citenamefont {Paxton},
  \citenamefont {Kistler}, \citenamefont {Olmeda}, \citenamefont {Sen},
  \citenamefont {Angelo}, \citenamefont {Cao}, \citenamefont {Mallouk},
  \citenamefont {Lammert},\ and\ \citenamefont {Crespi}}]{paxton04}%
  \BibitemOpen
  \bibfield  {author} {\bibinfo {author} {\bibfnamefont {W.~F.}\ \bibnamefont
  {Paxton}}, \bibinfo {author} {\bibfnamefont {K.~C.}\ \bibnamefont {Kistler}},
  \bibinfo {author} {\bibfnamefont {C.~C.}\ \bibnamefont {Olmeda}}, \bibinfo
  {author} {\bibfnamefont {A.}~\bibnamefont {Sen}}, \bibinfo {author}
  {\bibfnamefont {S.~K.~S.}\ \bibnamefont {Angelo}}, \bibinfo {author}
  {\bibfnamefont {Y.~Y.}\ \bibnamefont {Cao}}, \bibinfo {author} {\bibfnamefont
  {T.~E.}\ \bibnamefont {Mallouk}}, \bibinfo {author} {\bibfnamefont {P.~E.}\
  \bibnamefont {Lammert}}, \ and\ \bibinfo {author} {\bibfnamefont {V.~H.}\
  \bibnamefont {Crespi}},\ }\bibfield  {title} {\enquote {\bibinfo {title}
  {Catalytic nanomotors: {Autonomous} movement of striped nanorods},}\
  }\href@noop {} {\bibfield  {journal} {\bibinfo  {journal} {J. Am. Chem.
  Soc.}\ }\textbf {\bibinfo {volume} {126}},\ \bibinfo {pages} {13424--13431}
  (\bibinfo {year} {2004})}\BibitemShut {NoStop}%
\bibitem [{\citenamefont {Golestanian}, \citenamefont {Liverpool},\ and\
  \citenamefont {Ajdari}(2005)}]{golestanian05}%
  \BibitemOpen
  \bibfield  {author} {\bibinfo {author} {\bibfnamefont {R.}~\bibnamefont
  {Golestanian}}, \bibinfo {author} {\bibfnamefont {T.~B.}\ \bibnamefont
  {Liverpool}}, \ and\ \bibinfo {author} {\bibfnamefont {A.}~\bibnamefont
  {Ajdari}},\ }\bibfield  {title} {\enquote {\bibinfo {title} {Propulsion of a
  molecular machine by asymmetric distribution of reaction products},}\
  }\href@noop {} {\bibfield  {journal} {\bibinfo  {journal} {Phys. Rev. Lett.}\
  }\textbf {\bibinfo {volume} {94}},\ \bibinfo {pages} {220801} (\bibinfo
  {year} {2005})}\BibitemShut {NoStop}%
\bibitem [{\citenamefont {Howse}\ \emph {et~al.}(2007)\citenamefont {Howse},
  \citenamefont {Jones}, \citenamefont {Ryan}, \citenamefont {Gough},
  \citenamefont {Vafabakhsh},\ and\ \citenamefont {Golestanian}}]{howse07}%
  \BibitemOpen
  \bibfield  {author} {\bibinfo {author} {\bibfnamefont {J.~R.}\ \bibnamefont
  {Howse}}, \bibinfo {author} {\bibfnamefont {R.~A.~L.}\ \bibnamefont {Jones}},
  \bibinfo {author} {\bibfnamefont {A.~J.}\ \bibnamefont {Ryan}}, \bibinfo
  {author} {\bibfnamefont {T.}~\bibnamefont {Gough}}, \bibinfo {author}
  {\bibfnamefont {R.}~\bibnamefont {Vafabakhsh}}, \ and\ \bibinfo {author}
  {\bibfnamefont {R.}~\bibnamefont {Golestanian}},\ }\bibfield  {title}
  {\enquote {\bibinfo {title} {Self-motile colloidal particles: {From} directed
  propulsion to random walk},}\ }\href@noop {} {\bibfield  {journal} {\bibinfo
  {journal} {Phys. Rev. Lett.}\ }\textbf {\bibinfo {volume} {99}},\ \bibinfo
  {pages} {048102} (\bibinfo {year} {2007})}\BibitemShut {NoStop}%
\bibitem [{\citenamefont {Golestanian}, \citenamefont {Liverpool},\ and\
  \citenamefont {Ajdari}(2007)}]{golestanian07}%
  \BibitemOpen
  \bibfield  {author} {\bibinfo {author} {\bibfnamefont {R.}~\bibnamefont
  {Golestanian}}, \bibinfo {author} {\bibfnamefont {T.~B.}\ \bibnamefont
  {Liverpool}}, \ and\ \bibinfo {author} {\bibfnamefont {A.}~\bibnamefont
  {Ajdari}},\ }\bibfield  {title} {\enquote {\bibinfo {title} {Designing
  phoretic micro- and nano-swimmers},}\ }\href@noop {} {\bibfield  {journal}
  {\bibinfo  {journal} {New J. Phys.}\ }\textbf {\bibinfo {volume} {9}},\
  \bibinfo {pages} {126} (\bibinfo {year} {2007})}\BibitemShut {NoStop}%
\bibitem [{\citenamefont {Wang}(2009)}]{chemical1}%
  \BibitemOpen
  \bibfield  {author} {\bibinfo {author} {\bibfnamefont {J.}~\bibnamefont
  {Wang}},\ }\bibfield  {title} {\enquote {\bibinfo {title} {Can man-made
  nanomachines compete with nature biomotors?}}\ }\href {\doibase
  10.1021/nn800829k} {\bibfield  {journal} {\bibinfo  {journal} {ACS Nano}\
  }\textbf {\bibinfo {volume} {3}},\ \bibinfo {pages} {4--9} (\bibinfo {year}
  {2009})}\BibitemShut {NoStop}%
\bibitem [{\citenamefont {Mallouk}\ and\ \citenamefont
  {Sen}(2009)}]{chemical2}%
  \BibitemOpen
  \bibfield  {author} {\bibinfo {author} {\bibfnamefont {T.~E.}\ \bibnamefont
  {Mallouk}}\ and\ \bibinfo {author} {\bibfnamefont {A.}~\bibnamefont {Sen}},\
  }\bibfield  {title} {\enquote {\bibinfo {title} {Powering nanorobots},}\
  }\href {\doibase 10.1038/scientificamerican0509-72} {\bibfield  {journal}
  {\bibinfo  {journal} {Sci. Am.}\ }\textbf {\bibinfo {volume} {300}},\
  \bibinfo {pages} {72--77} (\bibinfo {year} {2009})}\BibitemShut {NoStop}%
\bibitem [{\citenamefont {Mirkovic}, \citenamefont {Nicole S.~Zacharia},\ and\
  \citenamefont {Ozin}(2010)}]{chemical3}%
  \BibitemOpen
  \bibfield  {author} {\bibinfo {author} {\bibfnamefont {T.}~\bibnamefont
  {Mirkovic}}, \bibinfo {author} {\bibfnamefont {G.~D.~S.}\ \bibnamefont
  {Nicole S.~Zacharia}}, \ and\ \bibinfo {author} {\bibfnamefont {G.~A.}\
  \bibnamefont {Ozin}},\ }\bibfield  {title} {\enquote {\bibinfo {title}
  {Nanolocomotion---catalytic nanomotors and nanorotors},}\ }\href {\doibase
  10.1021/nn100669h} {\bibfield  {journal} {\bibinfo  {journal} {ACS Nano}\
  }\textbf {\bibinfo {volume} {4}},\ \bibinfo {pages} {1782--1789} (\bibinfo
  {year} {2010})}\BibitemShut {NoStop}%
\bibitem [{\citenamefont {Tierno}, \citenamefont {Guell},\ and\ \citenamefont
  {Sagues}(2010)}]{surface1}%
  \BibitemOpen
  \bibfield  {author} {\bibinfo {author} {\bibfnamefont {P.}~\bibnamefont
  {Tierno}}, \bibinfo {author} {\bibfnamefont {O.}~\bibnamefont {Guell}}, \
  and\ \bibinfo {author} {\bibfnamefont {F.}~\bibnamefont {Sagues}},\
  }\bibfield  {title} {\enquote {\bibinfo {title} {Controlled propulsion in
  viscous fluids of magnetically actuated colloidal doublets},}\ }\href
  {\doibase 10.1103/PhysRevE.81.011402} {\bibfield  {journal} {\bibinfo
  {journal} {Phys. Rev. E}\ }\textbf {\bibinfo {volume} {81}},\ \bibinfo
  {pages} {011402} (\bibinfo {year} {2010})}\BibitemShut {NoStop}%
\bibitem [{\citenamefont {Sing}\ \emph {et~al.}(2010)\citenamefont {Sing},
  \citenamefont {Schmid}, \citenamefont {Schneider}, \citenamefont {Franke},\
  and\ \citenamefont {Alexander-Katz}}]{surface2}%
  \BibitemOpen
  \bibfield  {author} {\bibinfo {author} {\bibfnamefont {C.~E.}\ \bibnamefont
  {Sing}}, \bibinfo {author} {\bibfnamefont {L.}~\bibnamefont {Schmid}},
  \bibinfo {author} {\bibfnamefont {M.~F.}\ \bibnamefont {Schneider}}, \bibinfo
  {author} {\bibfnamefont {T.}~\bibnamefont {Franke}}, \ and\ \bibinfo {author}
  {\bibfnamefont {A.}~\bibnamefont {Alexander-Katz}},\ }\bibfield  {title}
  {\enquote {\bibinfo {title} {Controlled surface-induced flows from the motion
  of self-assembled colloidal walkers},}\ }\href {\doibase
  10.1073/pnas.0906489107} {\bibfield  {journal} {\bibinfo  {journal} {Proc.
  Natl. Acad. Sci. USA}\ }\textbf {\bibinfo {volume} {107}},\ \bibinfo {pages}
  {535} (\bibinfo {year} {2010})}\BibitemShut {NoStop}%
\bibitem [{\citenamefont {Zhang}\ \emph {et~al.}(2010)\citenamefont {Zhang},
  \citenamefont {Petit}, \citenamefont {Lu}, \citenamefont {Kratochvil},
  \citenamefont {Peyer}, \citenamefont {Ryan~Pei},\ and\ \citenamefont
  {Nelson}}]{surface3}%
  \BibitemOpen
  \bibfield  {author} {\bibinfo {author} {\bibfnamefont {L.}~\bibnamefont
  {Zhang}}, \bibinfo {author} {\bibfnamefont {T.}~\bibnamefont {Petit}},
  \bibinfo {author} {\bibfnamefont {Y.}~\bibnamefont {Lu}}, \bibinfo {author}
  {\bibfnamefont {B.~E.}\ \bibnamefont {Kratochvil}}, \bibinfo {author}
  {\bibfnamefont {K.~E.}\ \bibnamefont {Peyer}}, \bibinfo {author}
  {\bibfnamefont {J.~L.}\ \bibnamefont {Ryan~Pei}}, \ and\ \bibinfo {author}
  {\bibfnamefont {B.~J.}\ \bibnamefont {Nelson}},\ }\bibfield  {title}
  {\enquote {\bibinfo {title} {Controlled propulsion and cargo transport of
  rotating nickel nanowires near a patterned solid surface},}\ }\href@noop {}
  {\bibfield  {journal} {\bibinfo  {journal} {ACS Nano}\ }\textbf {\bibinfo
  {volume} {4}},\ \bibinfo {pages} {6228} (\bibinfo {year} {2010})}\BibitemShut
  {NoStop}%
\bibitem [{\citenamefont {Berg}(2003)}]{ecoli}%
  \BibitemOpen
  \bibfield  {author} {\bibinfo {author} {\bibfnamefont {H.~C.}\ \bibnamefont
  {Berg}},\ }\href@noop {} {\emph {\bibinfo {title} {E. Coli in Motion}}}\
  (\bibinfo  {publisher} {Springer},\ \bibinfo {address} {{\color{black}New
  York}},\ \bibinfo {year} {2003})\BibitemShut {NoStop}%
\bibitem [{\citenamefont {Dreyfus}\ \emph {et~al.}(2005)\citenamefont
  {Dreyfus}, \citenamefont {Baudry}, \citenamefont {Roper}, \citenamefont
  {Fermigier}, \citenamefont {Stone},\ and\ \citenamefont
  {Bibette}}]{flexible1}%
  \BibitemOpen
  \bibfield  {author} {\bibinfo {author} {\bibfnamefont {R.}~\bibnamefont
  {Dreyfus}}, \bibinfo {author} {\bibfnamefont {J.}~\bibnamefont {Baudry}},
  \bibinfo {author} {\bibfnamefont {M.~L.}\ \bibnamefont {Roper}}, \bibinfo
  {author} {\bibfnamefont {M.}~\bibnamefont {Fermigier}}, \bibinfo {author}
  {\bibfnamefont {H.~A.}\ \bibnamefont {Stone}}, \ and\ \bibinfo {author}
  {\bibfnamefont {J.}~\bibnamefont {Bibette}},\ }\bibfield  {title} {\enquote
  {\bibinfo {title} {Microscopic artificial swimmers},}\ }\href@noop {}
  {\bibfield  {journal} {\bibinfo  {journal} {Nature}\ }\textbf {\bibinfo
  {volume} {437}},\ \bibinfo {pages} {862--865} (\bibinfo {year}
  {2005})}\BibitemShut {NoStop}%
\bibitem [{\citenamefont {Gao}\ \emph {et~al.}(2010)\citenamefont {Gao},
  \citenamefont {Sattayasamitsathit}, \citenamefont {Manesh}, \citenamefont
  {Weihs},\ and\ \citenamefont {Wang}}]{flexible2}%
  \BibitemOpen
  \bibfield  {author} {\bibinfo {author} {\bibfnamefont {W.}~\bibnamefont
  {Gao}}, \bibinfo {author} {\bibfnamefont {S.}~\bibnamefont
  {Sattayasamitsathit}}, \bibinfo {author} {\bibfnamefont {K.~M.}\ \bibnamefont
  {Manesh}}, \bibinfo {author} {\bibfnamefont {D.}~\bibnamefont {Weihs}}, \
  and\ \bibinfo {author} {\bibfnamefont {J.}~\bibnamefont {Wang}},\ }\bibfield
  {title} {\enquote {\bibinfo {title} {Magnetically powered flexible metal
  nanowire motors},}\ }\href@noop {} {\bibfield  {journal} {\bibinfo  {journal}
  {J. Am. Chem. Soc}\ }\textbf {\bibinfo {volume} {132}},\ \bibinfo {pages}
  {14403} (\bibinfo {year} {2010})}\BibitemShut {NoStop}%
\bibitem [{\citenamefont {Pak}\ \emph {et~al.}(2011)\citenamefont {Pak},
  \citenamefont {Gao}, \citenamefont {Wang},\ and\ \citenamefont
  {Lauga}}]{OnshunSoftMatter}%
  \BibitemOpen
  \bibfield  {author} {\bibinfo {author} {\bibfnamefont {O.~S.}\ \bibnamefont
  {Pak}}, \bibinfo {author} {\bibfnamefont {W.}~\bibnamefont {Gao}}, \bibinfo
  {author} {\bibfnamefont {J.}~\bibnamefont {Wang}}, \ and\ \bibinfo {author}
  {\bibfnamefont {E.}~\bibnamefont {Lauga}},\ }\bibfield  {title} {\enquote
  {\bibinfo {title} {High-speed propulsion of flexible nanowire motors: Theory
  and experiments},}\ }\href@noop {} {\bibfield  {journal} {\bibinfo  {journal}
  {Soft Matter}\ }\textbf {\bibinfo {volume} {7}},\ \bibinfo {pages}
  {8169--8181} (\bibinfo {year} {2011})}\BibitemShut {NoStop}%
\bibitem [{\citenamefont {Zhang}, \citenamefont {Peyer},\ and\ \citenamefont
  {Nelson}(2010)}]{NelsonRev}%
  \BibitemOpen
  \bibfield  {author} {\bibinfo {author} {\bibfnamefont {L.}~\bibnamefont
  {Zhang}}, \bibinfo {author} {\bibfnamefont {K.~E.}\ \bibnamefont {Peyer}}, \
  and\ \bibinfo {author} {\bibfnamefont {B.~J.}\ \bibnamefont {Nelson}},\
  }\bibfield  {title} {\enquote {\bibinfo {title} {Artificial bacteria flagella
  for micromanipulation},}\ }\href@noop {} {\bibfield  {journal} {\bibinfo
  {journal} {Lab on a Chip}\ }\textbf {\bibinfo {volume} {10}},\ \bibinfo
  {pages} {2203--2215} (\bibinfo {year} {2010})}\BibitemShut {NoStop}%
\bibitem [{\citenamefont {Ghosh}\ and\ \citenamefont
  {Fischer}(2009)}]{GhoshFischer09}%
  \BibitemOpen
  \bibfield  {author} {\bibinfo {author} {\bibfnamefont {A.}~\bibnamefont
  {Ghosh}}\ and\ \bibinfo {author} {\bibfnamefont {P.}~\bibnamefont
  {Fischer}},\ }\bibfield  {title} {\enquote {\bibinfo {title} {Controlled
  propulsion of artificial magnetic nanostructured propellers},}\ }\href
  {\doibase 10.1021/nl900186w} {\bibfield  {journal} {\bibinfo  {journal} {Nano
  Lett.}\ }\textbf {\bibinfo {volume} {9}},\ \bibinfo {pages} {2243--2245}
  (\bibinfo {year} {2009})}\BibitemShut {NoStop}%
\bibitem [{\citenamefont {Ghosh}\ \emph {et~al.}(2012)\citenamefont {Ghosh},
  \citenamefont {Paria}, \citenamefont {Singh}, \citenamefont {Venugopalan},\
  and\ \citenamefont {Ghosh}}]{ghosh12}%
  \BibitemOpen
  \bibfield  {author} {\bibinfo {author} {\bibfnamefont {A.}~\bibnamefont
  {Ghosh}}, \bibinfo {author} {\bibfnamefont {D.}~\bibnamefont {Paria}},
  \bibinfo {author} {\bibfnamefont {H.~J.}\ \bibnamefont {Singh}}, \bibinfo
  {author} {\bibfnamefont {P.}~\bibnamefont {Venugopalan}}, \ and\ \bibinfo
  {author} {\bibfnamefont {A.}~\bibnamefont {Ghosh}},\ }\bibfield  {title}
  {\enquote {\bibinfo {title} {Dynamical configurations and bistability of
  helical nanostructures under external torque},}\ }\href@noop {} {\bibfield
  {journal} {\bibinfo  {journal} {Phys. Rev. E}\ }\textbf {\bibinfo {volume}
  {86}},\ \bibinfo {pages} {031401} (\bibinfo {year} {2012})}\BibitemShut
  {NoStop}%
\bibitem [{\citenamefont {Peyer}, \citenamefont {Li~Zhang},\ and\ \citenamefont
  {Nelson}(2010)}]{NelsonNonideal}%
  \BibitemOpen
  \bibfield  {author} {\bibinfo {author} {\bibfnamefont {K.~E.}\ \bibnamefont
  {Peyer}}, \bibinfo {author} {\bibfnamefont {B.~E.~K.}\ \bibnamefont
  {Li~Zhang}}, \ and\ \bibinfo {author} {\bibfnamefont {B.~J.}\ \bibnamefont
  {Nelson}},\ }\bibfield  {title} {\enquote {\bibinfo {title} {Non-ideal
  swimming of artificial bacterial flagella near a surface},}\ }in\ \href@noop
  {} {\emph {\bibinfo {booktitle} {Proceedings of the 2010 IEEE International
  Conference on Robotics and Automation}}}\ (\bibinfo {address} {Anchorage,
  AK},\ \bibinfo {year} {2010})\ pp.\ \bibinfo {pages} {96--101}\BibitemShut
  {NoStop}%
\bibitem [{\citenamefont {Tottori}\ \emph {et~al.}(2012)\citenamefont
  {Tottori}, \citenamefont {Zhang}, \citenamefont {Qiu}, \citenamefont
  {Krawczyk}, \citenamefont {Franco-Obregon},\ and\ \citenamefont
  {Nelson}}]{NelsonCargo}%
  \BibitemOpen
  \bibfield  {author} {\bibinfo {author} {\bibfnamefont {S.}~\bibnamefont
  {Tottori}}, \bibinfo {author} {\bibfnamefont {L.}~\bibnamefont {Zhang}},
  \bibinfo {author} {\bibfnamefont {F.}~\bibnamefont {Qiu}}, \bibinfo {author}
  {\bibfnamefont {K.~K.}\ \bibnamefont {Krawczyk}}, \bibinfo {author}
  {\bibfnamefont {A.}~\bibnamefont {Franco-Obregon}}, \ and\ \bibinfo {author}
  {\bibfnamefont {B.~J.}\ \bibnamefont {Nelson}},\ }\bibfield  {title}
  {\enquote {\bibinfo {title} {Magnetic helical micromachines: Fabrication,
  controlled swimming, and cargo transport},}\ }\href {\doibase
  10.1002/adma.201103818} {\bibfield  {journal} {\bibinfo  {journal} {Advanced
  Material}\ }\textbf {\bibinfo {volume} {24}} (\bibinfo {year} {2012}),\
  10.1002/adma.201103818}\BibitemShut {NoStop}%
\bibitem [{\citenamefont {Gray}\ and\ \citenamefont {Hancock}(1955)}]{RFT}%
  \BibitemOpen
  \bibfield  {author} {\bibinfo {author} {\bibfnamefont {J.}~\bibnamefont
  {Gray}}\ and\ \bibinfo {author} {\bibfnamefont {G.~J.}\ \bibnamefont
  {Hancock}},\ }\bibfield  {title} {\enquote {\bibinfo {title} {The propulsion
  of sea-urchin spermatozoa},}\ }\href@noop {} {\bibfield  {journal} {\bibinfo
  {journal} {J. Exp. Biol.}\ }\textbf {\bibinfo {volume} {32}},\ \bibinfo
  {pages} {802--814} (\bibinfo {year} {1955})}\BibitemShut {NoStop}%
\bibitem [{\citenamefont {G.Cox}(1970)}]{Cox1970}%
  \BibitemOpen
  \bibfield  {author} {\bibinfo {author} {\bibfnamefont {R.}~\bibnamefont
  {G.Cox}},\ }\bibfield  {title} {\enquote {\bibinfo {title} {The motion of
  long slender bodies in a viscous fluid},}\ }\href@noop {} {\bibfield
  {journal} {\bibinfo  {journal} {J. Fluid Mech.}\ }\textbf {\bibinfo {volume}
  {44}},\ \bibinfo {pages} {791--810} (\bibinfo {year} {1970})}\BibitemShut
  {NoStop}%
\bibitem [{\citenamefont {Batchelor}(1970)}]{Batchelor1970}%
  \BibitemOpen
  \bibfield  {author} {\bibinfo {author} {\bibfnamefont {G.~K.}\ \bibnamefont
  {Batchelor}},\ }\bibfield  {title} {\enquote {\bibinfo {title} {Slender-body
  theory for particles of arbitrary cross-section in stokes flow},}\
  }\href@noop {} {\bibfield  {journal} {\bibinfo  {journal} {J. Fluid Mech.}\
  }\textbf {\bibinfo {volume} {44}},\ \bibinfo {pages} {419--440} (\bibinfo
  {year} {1970})}\BibitemShut {NoStop}%
\bibitem [{\citenamefont {Keller}\ and\ \citenamefont
  {Rubinow}(1976)}]{keller76-jfm}%
  \BibitemOpen
  \bibfield  {author} {\bibinfo {author} {\bibfnamefont {J.~B.}\ \bibnamefont
  {Keller}}\ and\ \bibinfo {author} {\bibfnamefont {S.~I.}\ \bibnamefont
  {Rubinow}},\ }\bibfield  {title} {\enquote {\bibinfo {title} {Slender body
  theory for slow viscous flow},}\ }\href@noop {} {\bibfield  {journal}
  {\bibinfo  {journal} {J. Fluid Mech.}\ }\textbf {\bibinfo {volume} {75}},\
  \bibinfo {pages} {705--714} (\bibinfo {year} {1976})}\BibitemShut {NoStop}%
\bibitem [{\citenamefont {Johnson}(1980)}]{johnson80}%
  \BibitemOpen
  \bibfield  {author} {\bibinfo {author} {\bibfnamefont {R.~E.}\ \bibnamefont
  {Johnson}},\ }\bibfield  {title} {\enquote {\bibinfo {title} {An improved
  slender body theory for {S}tokes flow},}\ }\href@noop {} {\bibfield
  {journal} {\bibinfo  {journal} {J. Fluid Mech.}\ }\textbf {\bibinfo {volume}
  {99}},\ \bibinfo {pages} {411--431} (\bibinfo {year} {1980})}\BibitemShut
  {NoStop}%
\bibitem [{\citenamefont {Pak}, \citenamefont {Spagnolie},\ and\ \citenamefont
  {Lauga}(2012)}]{Pak2012}%
  \BibitemOpen
  \bibfield  {author} {\bibinfo {author} {\bibfnamefont {O.~S.}\ \bibnamefont
  {Pak}}, \bibinfo {author} {\bibfnamefont {S.~E.}\ \bibnamefont {Spagnolie}},
  \ and\ \bibinfo {author} {\bibfnamefont {E.}~\bibnamefont {Lauga}},\
  }\bibfield  {title} {\enquote {\bibinfo {title} {{\color{black}Hydrodynamics
  of the double-wave structure of insect spermatozoa flagella}},}\ }\href@noop
  {} {\bibfield  {journal} {\bibinfo  {journal} {{\color{black}J.R. Soc.
  Interface}}\ }\textbf {\bibinfo {volume} {{\color{black}9}}},\ \bibinfo
  {pages} {1908--1924} (\bibinfo {year} {{\color{black}2012}})}\BibitemShut
  {NoStop}%
\bibitem [{\citenamefont {Jung}\ \emph {et~al.}(2007)\citenamefont {Jung},
  \citenamefont {Mareck}, \citenamefont {Fauci},\ and\ \citenamefont
  {Shelley}}]{Jung07}%
  \BibitemOpen
  \bibfield  {author} {\bibinfo {author} {\bibfnamefont {S.}~\bibnamefont
  {Jung}}, \bibinfo {author} {\bibfnamefont {K.}~\bibnamefont {Mareck}},
  \bibinfo {author} {\bibfnamefont {L.}~\bibnamefont {Fauci}}, \ and\ \bibinfo
  {author} {\bibfnamefont {M.~J.}\ \bibnamefont {Shelley}},\ }\bibfield
  {title} {\enquote {\bibinfo {title} {{\color{black}Rotational dynamics of a
  superhelix towed in a Stokes fluid}},}\ }\href@noop {} {\bibfield  {journal}
  {\bibinfo  {journal} {{\color{black}Phys. Fluids}}\ }\textbf {\bibinfo
  {volume} {19}} (\bibinfo {year} {{\color{black}2007}})}\BibitemShut {NoStop}%
\bibitem [{\citenamefont {Wolgemuth}, \citenamefont {Powers},\ and\
  \citenamefont {Goldstein}(2000)}]{Spinning}%
  \BibitemOpen
  \bibfield  {author} {\bibinfo {author} {\bibfnamefont {C.~W.}\ \bibnamefont
  {Wolgemuth}}, \bibinfo {author} {\bibfnamefont {T.~R.}\ \bibnamefont
  {Powers}}, \ and\ \bibinfo {author} {\bibfnamefont {R.~E.}\ \bibnamefont
  {Goldstein}},\ }\bibfield  {title} {\enquote {\bibinfo {title} {Twirling and
  whirling: Viscous dynamics of rotating elastic filaments},}\ }\href@noop {}
  {\bibfield  {journal} {\bibinfo  {journal} {Phys. Rev. Lett.}\ }\textbf
  {\bibinfo {volume} {84}} (\bibinfo {year} {2000})}\BibitemShut {NoStop}%
\end{thebibliography}%

\end{document}